\documentclass{emulateapj_may2010}

\usepackage{xspace}
\usepackage{graphicx}
\usepackage{natbib}
\usepackage{apjfonts}
\def\errtwo#1#2#3{$#1^{+#2}_{-#3}$}

\newcommand\msun{{\rm M_\odot}}

\newcommand\cyg{{Cyg~X-1}\xspace}

\newcommand\zth{$0^{\rm th}$}
\newcommand\fst{$1^{\rm st}$}

\newcommand\asm{{ASM}\xspace}

\newcommand\beppo{\textsl{BeppoSAX}\xspace}
\newcommand\chandra{\textsl{Chandra}\xspace}

\newcommand\gso{{GSO}\xspace}
\newcommand\heg{{HEG}\xspace}
\newcommand\hetg{{HETG}\xspace}
\newcommand\hexte{{HEXTE}\xspace}
\newcommand\hxd{{HXD}\xspace}
\newcommand\isis{{\tt ISIS}\xspace}
\newcommand\meg{{MEG}\xspace}

\newcommand\pca{{PCA}\xspace}
\newcommand\pin{{PIN}\xspace}

\newcommand\rxte{\textsl{RXTE}\xspace}
\newcommand\ryle{\textsl{Ryle}\xspace}

\newcommand\suzaku{\textsl{Suzaku}\xspace}
\newcommand\xis{{XIS}\xspace}

\newcommand\xspec{{\tt XSPEC}\xspace}

\newcommand\aproxgt{\mathrel{%
     \rlap{\raise 0.511ex \hbox{$>$}}{\lower 0.511ex \hbox{$\sim$}}}}
\newcommand\aproxlt{\mathrel{%
     \rlap{\raise 0.511ex \hbox{$<$}}{\lower 0.511ex \hbox{$\sim$}}}}

\slugcomment{Received 2010 February 18; ~~accepted 2010 November 24}

\shorttitle{Observations of the Cygnus X-1 Hard State}
\shortauthors{Nowak et al.}

\begin{document}

\title{Corona, Jet, and Relativistic Line Models for Suzaku/RXTE/Chandra-HETG \\
Observations of the Cygnus X-1 Hard State}

\author{Michael A. Nowak\altaffilmark{1}, Manfred
  Hanke\altaffilmark{2}, Sarah N. Trowbridge\altaffilmark{1}, Sera
  B. Markoff\altaffilmark{3}, J\"orn Wilms\altaffilmark{2}, \\ Katja
  Pottschmidt\altaffilmark{4}, Paolo Coppi\altaffilmark{5}, Dipankar
  Maitra\altaffilmark{3,6}, John E. Davis\altaffilmark{1}, Frank
  Tramper\altaffilmark{3}} \altaffiltext{1}{Massachusetts Institute of
  Technology, Kavli Institute for Astrophysics, Cambridge, MA 02139,
  USA; mnowak,davis@space.mit.edu, saraht@mit.edu}
\altaffiltext{2}{Dr.~Karl Remeis-Sternwarte and Erlangen Centre for
  Astroparticle Physics, Universit\"at Erlangen-N\"urnberg,
  Sternwartstr.~7, 96049 Bamberg, Germany; Manfred.Hanke,
  joern.wilms@sternwarte.uni-erlangen.de}
\altaffiltext{3}{Astronomical Institute ``Anton Pannekoek'' University
  of Amsterdam; s.b.markoff@uva.nl} \altaffiltext{4}{CRESST, UMBC, and
  NASA Goddard Space Flight Center, Greenbelt, MD 20771;
  katja@milkyway.gsfc.nasa.gov} \altaffiltext{5}{Yale University, New
  Haven, CT; paolo.coppi@yale.edu} \altaffiltext{6}{current address:
  Dept. of Astronomy, Univ. of Michigan, 500 Church St., Ann Arbor, MI
  48109-1042; dmaitra@umich.edu}

\begin{abstract}

Using \suzaku and the \textsl{Rossi X-ray Timing Explorer} (\rxte), we
have conducted a series of four simultaneous observations of the
galactic black hole candidate \cyg in what were historically faint and
spectrally hard ``low states''.  Additionally, all of these
observations occurred near superior conjunction with our line of sight
to the X-ray source passing through the dense phases of the ``focused
wind'' from the mass donating secondary.  One of our observations was
also simultaneous with observations by the \chandra-High Energy
Transmission Grating (\hetg).  These latter spectra are crucial for
{revealing} the ionized absorption due to the secondary's focused
wind. Such absorption is present and must be accounted for in all four
spectra.  These simultaneous data give an unprecedented view of the
0.8--300\,keV spectrum of \cyg, and hence bear upon both corona and
X-ray emitting jet models of black hole hard states. Three models fit
the spectra well: coronae with thermal or mixed thermal/non-thermal
electron populations, and jets. All three models require a soft
component that we fit with a low temperature disk spectrum with an
inner radius of only a few tens of $GM/c^2$. All three models also
agree that the known spectral break at 10\,keV is not solely due to
the presence of reflection, but each gives a different underlying
explanation for the augmentation of this break. Thus whereas all three
models require that there is a relativistically broadened Fe line, the
strength and inner radius of such a line is dependent upon the
specific model, {thus making premature line-based estimates of the
  black hole spin in the Cyg X-1 system}. We look at the relativistic
line in detail, accounting for the narrow Fe emission and ionized
absorption detected by \hetg. Although the specific relativistic
parameters of the line are continuum-dependent, none of the broad line
fits allow for an inner disk radius that is $>40$\,$GM/c^2$.

\end{abstract}


\section{Introduction}\label{sec:intro}

\setcounter{footnote}{0}

There is currently significant debate as to the physical mechanisms
responsible for the continuum of X-ray spectrally ``hard states'' of
black hole candidates (BHC) accreting in binaries.  This debate ranges
from the broader issue of whether or not there is a significant
contribution to the X-ray band from an outflow or jet, to more
narrowly focused issues within given classes of models.  For instance,
the hard X-ray emission has traditionally been attributed to a
Comptonizing thermal corona \citep[][and references
  therein]{eardley:75a,shapiro:76a,sunyaev:79a,dove:98a}.  These
earlier works generally favor a scenario where the corona lies central
to a truncated outer thin disk.  However, if the corona is driven
outwards by radiative pressure \citep[e.g.,][] {beloborodov:99a} could
it instead overlay the inner disk?  Can the optically thick,
geometrically thin disk extend inward nearly to the innermost stable
circular orbit \citep{miller:06b}?  Is this disk cold (peak
temperatures of a few hundred eV), or can it instead be hot (near a
keV, i.e., \citealt{wilms:06a})? Is the hard state corona comprised
primarily of electrons with a thermal population \citep{poutanen:09a},
or can it have a substantial contribution from a non-thermal electron
population \citep{ibragimov:05a}? Does the bulk motion of the flow
play a role in Comptonizing the spectrum
\citep{shaposhnikov:06a,laurent:07a}?  Alternatively could the X-rays be
comprised of a combination of direct synchrotron and synchrotron
self-Compton (SSC) emission from a jet
\citep{markoff:05a,maitra:09a}?  Contributing to the debate, however,
is the fact that many of the above cited models, especially when
considering solely \textsl{Rossi X-ray Timing Explorer} (\rxte) data
in the 3--200\,keV range (or even narrower energy ranges), describe
the hard state spectra nearly equally well.

To study these issues, over the past decade we have been using a
series of pointed, approximately bi-weekly, \rxte observations of the
BHC \cyg performed simultaneously with 15\,GHz radio observations by
the Ryle telescope.  \cyg holds much promise for exploring the current
range of questions listed above owing to its persistently bright
X-ray flux (including both the hard and soft states, it varies between
200--600\,mCrab in the 1.2--12 keV band covered by the \rxte-All Sky
Monitor) and its correlated radio/X-ray spectra \citep[see][and
  references therein]{wilms:06a}. Extended radio emission even has
been imaged in Cyg X-1 \citep{stirling:01a}.  This campaign has
already provided the spectra for some of the Comptonization models
\citep{wilms:06a}, and jet models \citep{markoff:05a} discussed above.
Additionally, these data have been used to to study the correlation of
the X-ray and radio spectral properties on both long time scales
\citep{pottschmidt:00a,pottschmidt:02a,wilms:06a}, and short time
scales (\citealt{gleissner:04a}, \citealt{wilms:07a},
\citealt{boeck:10a}, B\"ock et al., in prep.).  Thus they comprise a
strong data set for addressing not only the details of the X-ray
spectrum, but also the connection to the hard state jets which are
known to dominate the radio through near-infrared emission.
Simultaneous radio/X-ray flaring has also been detected in \cyg during
this extended campaign (\citealt{fender:06a}, \citealt{wilms:07a}),
lending support to the hypothesis of X-ray emission by the jet.

\rxte spectral data alone, however, do not allow us to break the
current existing ``theoretical degeneracy'' in the origin of the X-ray
spectrum.  The statistically best fits are in fact obtained with
purely empirical simple broken powerlaws with a break occurring
between 9--12\,keV, and an exponential cutoff occurring at >20\,keV,
to which a broad, $\approx 6.4$\,keV gaussian line is added
\citep{wilms:06a,nowak:05a}.  The latter component is likely
attributable to a relativistically broadened Fe K$\alpha$ line
\citep[][and references therein]{reynolds:03a}; however, its
parameters are dependent upon the assumed continuum model
\citep{wilms:06a}.  More physically motivated Comptonization and
outflow-dominated models \citep{markoff:05a,wilms:06a} can describe
the same spectra almost as well.  However they must introduce
additional, albeit plausible, physical components (e.g., relativistic
smearing) to recover the simple spectra of the broken powerlaw
description. Thus, there is some amount of ambiguity when correlating
detailed spectral features vs. continuum properties, e.g., disk
reflection vs. coronal compactness/hardness, as the detailed features
systematically depend upon the underlying broad-band continuum.

Despite these problems in finding a truly unique spectral model, when
considering multiple observations taken over a wide range of
luminosities and spectral hardnesses, spectral correlations arise that
are robust and persistent across a variety of these theoretical
characterizations \citep{wilms:06a}.  Using the broken powerlaw models
as a simple description of the X-ray spectra, we have found that when
the 2--10\,keV photon index $\Gamma_1 < 2.2$ there is a positive
correlation between X-ray and radio flux, whereas for $\Gamma_1 > 2.2$
there is a negative correlation between the X-ray and radio flux. We
therefore use the value of $\Gamma_1 \approx 2.2$ as the canonical
division between the spectrally ``hard state'' and the spectrally
``soft state'' \citep[see][]{remillard:06a}.  Additionally, as the
spectra become harder, exponential cutoffs tend to become less
significant \citep[see also][]{motta:09a}.

Our previous \rxte spectral studies of \cyg have been limited in two
respects: the low spectral resolution of \rxte ($E/\Delta E \approx 6$
at 6\,keV), and the inability to measure spectra at $\aproxlt
3$\,keV. In this work, we turn to a set of four \suzaku observations
that we performed simultaneously with our \rxte campaign to enhance
our \cyg studies in several crucial ways. First, \suzaku has large
effective area at soft X-ray energies. In this work, we consider
spectra down to 0.8\,keV, which allows for the possibility of
measuring the ``seed photon spectrum'' in Comptonization models
(\S\ref{sec:compton}), or judging the relative contribution of
synchrotron versus the disk radiation in jet models (\S\ref{sec:jet}).
\suzaku also has excellent resolution in the Fe K$\alpha$ line region
($E/\Delta E \approx 50$), which allows us to separate narrow from
relativistically broadened line features (\S\ref{sec:comp_line},
\ref{sec:rel_line}). Third, \suzaku measures the \cyg hard X-ray
spectrum up to $\approx 300$\,keV (\S\ref{sec:cutoff}), providing
further constraints on Comptonization and jet models.

\begin{deluxetable}{ccccc}  
\setlength{\tabcolsep}{0.03in} 
\tabletypesize{\footnotesize}    
\tablewidth{0pt} 
\tablecaption{Log of \cyg Observations \label{tab:lo}}
\tablehead{\colhead{Date}
          & \colhead{Spacecraft/ObsID}
          & \colhead{Instrument}                   
          & \multicolumn{2}{c}{Exposure}
          \\                               
          (yyyy-mm-dd) & & & \multicolumn{2}{c}{(ksec)}
         }
\startdata
    2006-10-30 & \suzaku/401059010  & \xis0--3 & 35.0\tablenotemark{a} & 32.1
\\
    \nodata & \nodata & \hxd-\pin & 27.7 & 24.9
\\
    \nodata & \nodata & \hxd-\gso & 27.7 & 25.8
\\
    \nodata & \rxte/80110-01-13 & \pca & 8.0\tablenotemark{b} & 8.0
\\
    \nodata & \nodata & \hexte-A & 3.1 & 3.1
\\
    \nodata & \nodata & \hexte-B & 2.5 & 2.5
\\
    2007-04-30 & \suzaku/402072010  & \xis0,1,3 & 34.0\tablenotemark{a} & 22.4
\\
    \nodata & \nodata & \hxd-\pin & 40.2 & 27.6
\\
    \nodata & \nodata & \hxd-\gso & 40.2 & 27.1
\\
    \nodata & \rxte/92090-01-16 & \pca & 14.6\tablenotemark{b} & 1.5
\\
    \nodata & \nodata & \hexte-A & 4.6\tablenotemark{c,d} & 0.3
\\
    \nodata & \nodata & \hexte-B & 3.0 & 0.4
\\
    2007-05-17 & \suzaku/402072020  & \xis0,1,3 & 22.3\tablenotemark{a} & 12.3
\\
    \nodata & \nodata & \hxd-\pin & 32.6 & 10.9
\\
    \nodata & \nodata & \hxd-\gso & 32.6 & 10.9 
\\
    \nodata & \rxte/92090-01-17 & \pca & 2.0\tablenotemark{b} & 0.0
\\
    \nodata & \nodata & \hexte-A & 1.9\tablenotemark{c,d} & 0.0
\\
    \nodata & \nodata & \hexte-B & 0.9 & 0.0
\\
    2008-04-19 & \suzaku/403065010  & \xis1,3 & 16.9\tablenotemark{a} & 7.2 
\\
    \nodata & \nodata & \xis0 & 34.0\tablenotemark{e} & 0.0
\\
    \nodata & \nodata & \hxd-\pin & 29.0 & 10.1 
\\
    \nodata & \nodata & \hxd-\gso & 29.0 & 10.1
\\
    \nodata & \rxte/93120-01-01 & \pca & 21.5\tablenotemark{b} & 7.2
\\
    \nodata & \nodata & \hexte-A & 17.9\tablenotemark{c} & 7.1
\\
    \nodata & \nodata & \hexte-B & 10.8 & 4.6
\\
    \nodata & \chandra/8525 & \hetg & 29.4\tablenotemark{f} & 11.1\tablenotemark{g}
\\
\enddata 
\tablecomments{Exposure times are after initial good time filtering
  {(left), and after color/intensity time filtering (right)}.}
\tablenotetext{a}{Summed exposure times for all listed \xis detectors.}
\tablenotetext{b}{Summed exposure intervals, not weighted by the 
                 fraction of operating Proportional Counter Units (PCU).}
\tablenotetext{c}{HEXTE-A cluster in fixed position (no rocking).}
\tablenotetext{d}{Evidence for HEXTE-A cluster exposure time anomaly.}
\tablenotetext{e}{\xis0 run in continuous readout mode.}
\tablenotetext{f}{Summed exposure times for all $1^{\rm st}$ order spectra.}
\tablenotetext{g}{Summed exposure times for $1^{\rm st}$ order \heg spectra.}

\end{deluxetable} 

The outline of this paper is as follows. In \S\ref{sec:data}, we
describe our full set of observations and our data reduction
procedures. Due to maintenance and upgrade of the Ryle radio telescope
during the construction of the Arcminute Microkelvin Imager (AMI), no
simultaneous radio measurements are available for these \suzaku/\rxte
observations. Instead, we use previous observations to estimate the
radio fluxes (\S\ref{sec:radio}). For one of our observations
simultaneous \chandra-\hetg data are available
(\S\ref{sec:data_chandra}). These \chandra data become crucial in all
of our analyses as they help elucidate the spectral variability
associated with the observed lightcurve behavior, as discussed in
\S\ref{sec:lightcurves}. We present simple phenomenological
descriptions of the spectra in \S\ref{sec:simple}, including a
description of the composite line profile (\S\ref{sec:comp_line}).
Comptonization and jet models are presented in \S\ref{sec:bb}, along
with further discussions of the implied relativistic lines
(\S\ref{sec:rel_line}). We summarize our findings in
\S\ref{sec:discuss}.

\section{Observations and Data Analysis}\label{sec:data}

Prior to 2009 April, there have been five \suzaku observations of
\cyg. The four most recent of these observations are discussed in this
work. The first \suzaku \cyg observation occurred in 2005 October
\citep{makishima:08a}. During this observation \cyg\ was in a
relatively bright hard state (the \rxte-\asm count rate was $\approx
30$\,cps), the spacecraft aimpoint was placed on the \suzaku-X-ray
Imaging Spectrometer (\xis) detectors, and \suzaku was run in a
data-mode with 1\,s integrations per CCD exposure frame. As discussed
by \citet[][see also the Appendix]{makishima:08a}, this long exposure
led to both telemetry dropouts and severe photon pileup on the
detectors.  As these issues require a more complex analysis
\citep{makishima:08a}, we do not consider these data further.

The four \suzaku observations discussed here occurred during times
when the \rxte-\asm flux ranged from 12--23\,cps, the spacecraft
pointing was set to the Hard X-ray Detector (\hxd; this slightly
reduces the flux on the \xis detectors), and the CCD exposure frame
integration times were set to $\approx 0.5$\,s. Thus, our observations
do not suffer from telemetry dropouts and they are less severely
affected by photon pileup. All four of these observations occurred
simultaneously with \rxte observations. The last of these observations
also occurred simultaneously with observations by every other
X-ray/soft gamma-ray instrument flying at that time (Hanke et al., in
prep.). Here we only consider the simultaneous \chandra-High Energy
Transmission Grating.  An observing log is presented in
Table~\ref{tab:lo}.

\subsection{Suzaku Analysis}\label{sec:data_suzaku}

The \suzaku data were reduced with tools from the {HEASOFT v6.8
package and calibration files dated 2009 September 25}. The instruments on
\suzaku \citep{mitsuda:07a} are the X-ray Imaging Spectrometer
\citep[\xis;][]{koyama:07a} CCD detector covering the $\approx
0.3$--10\,keV band, and the Hard X-ray Detector
\citep[\hxd;][]{takahashi:07a} comprised of the PIN diode detector
(\pin) covering the $\approx 10$--70\,keV band and the gadolinium
silicate crystal detector (\gso) covering the $\approx~60$--600\,keV
band. The \xis has four separate detectors, \xis0--3, with \xis1 being
a backside illuminated CCD. \xis2 was lost due to a micrometeor hit in
late 2006, and thus was available only for the first observation (see
Table \ref{tab:lo}). For our fourth observation, \xis0 was run in
continuous readout mode, which is not yet fully calibrated, so we do
not include these data in this work.

In preparing the \xis spectra, we first corrected each detector for
Charge Transfer Inefficiency (CTI) using the \texttt{xispi} tool, and
then reprocessed the data with \texttt{xselect} using the standard
\texttt{xisrepro} selection criteria. Due to thermal flexing of the
spacecraft, the attitude of the \suzaku spacecraft exhibits
variability over the course of the observations and therefore the
image of the source is not at a fixed position on the CCD. Standard
processing reduces this variability and improves the PSF image
\citep{uchiyama:08a}; however, with the standard tools it is not
possible yet to correct fully the blurring caused by the varying
attitude. As described in the Appendix, for bright sources a better
reconstruction of the attitude solution, and thus narrower PSF images,
can be obtained using the \texttt{aeattcor.sl} software. We produced
such an improved image and then used the \texttt{pile\_estimate.sl}
\texttt{S-Lang} script described in the Appendix to estimate the
degree of pileup in the observations. For the spectra described in
this work, using the time filter criteria described in
\S\ref{sec:lightcurves}, the center of the PSF images could be
affected by as much as a 35\% pileup fraction.  We therefore extracted
annular regions wherein we excised the $\approx 20''$ radius central
region.  The excised data accounted for approximately 1/3 of the
detected events.  The outer radii of our annular extraction regions
were limited by the 1/4 sub-array used in our CCD readout mode, and
thus were $\approx 2'$.  We estimate that the extracted events had
$<4\%$ mean residual pileup fraction.  All regions of the CCD are
dominated by source counts, therefore we did not extract nor use any
background spectra for the \xis observations.

Events in the \xis detectors were read out from either $3\times3$ or
$2\times2$ pixel islands. The \xis1 data were always in
$3\times3$-mode, whereas the other detectors had mixtures of
$3\times3$- and $2\times2$-mode. We created individual spectra and
response files for each detector and data mode combination. Spectra
were created for the time intervals described in
\S\ref{sec:lightcurves}, {with the resulting exposure times also
  being listed in Table~\ref{tab:lo}}.  Response matrices and
effective area files were created with the \texttt{xisrmfgen} and
\texttt{xissimarfgen} tools, respectively.

Although for each observation the individual \xis spectra were fit
separately, they were jointly grouped on a common grid such that they
had a minimum combined signal-to-noise ratio of 8 in each energy bin
(i.e., 64 total counts in each bin) and that the minimum number of
channels per energy bin was at least the half width half maximum of
the spectral resolution\footnote{The half width half maximum (HWHM)
  was determined by using the spectral responses to create fake
  spectra of delta function lines at discrete energies.  These fake
  spectra were then fit without response matrices to determine
  HWHM. In practice, this meant that the original spectra with 4096
  channels each were grouped, using the \texttt{ISIS} \texttt{group}
  function, by a minimum of 6, 8, 12, 14, 16, 18, 20, 22 channels
  starting at 0.5, 1, 2, 3, 4, 5, 6, and 7\,keV, respectively. For the
  most part, the spectral binnings were dominated by the HWHM criteria
  rather than the signal-to-noise criterion.}. To avoid regions of
poorly understood response, we only considered spectral energy ranges
0.8--1.72\,keV, 1.88--2.19\,keV, and 2.37--7.5\,keV\footnote{Outside
  of these energy ranges there are large disagreements among the fit
  residuals for the individual detectors and data modes. The excised
  regions between 1.72--2.37\,keV correspond to poorly calibrated Si
  and Ir features related to the detectors and mirrors.}.

For the \pin, we extracted spectra from the ``cleaned'' event files in
the \texttt{hxd/event\_cl} directories. The appropriate response and
background files were downloaded from the High Energy Astrophysics
Science Archive Research Center (HEASARC), specifically those from the
\texttt{pinxb\_ver2.0\_tuned} directory. Good time intervals (GTI)
were merged from our time selections
(\S\ref{sec:lightcurves}) and the combination of the GTI intervals
from the source and background event files. These intersected
intervals were then used to extract the \pin source and background
spectra. The source spectra exposure times were then corrected with
the \texttt{hxdtcorr} tool. The \pin spectra were grouped to have a
signal-to-noise ratio $\ge 10$ in each energy bin, and we considered
spectra between 12--70\,keV.

The \gso spectra were created starting with the ``unfiltered'' event
files. These data were first reprocessed with the \texttt{hxdtime},
\texttt{hxdpi}, and \texttt{hxdgrade} tools, following the
``\emph{Suzaku ABC Guide}''. The data were then filtered in
\texttt{xselect} with the standard criteria from the HEASARC provided
\texttt{gso\_mkf.sel} script. The background was downloaded from the
\texttt{gsonxb\_ver2.0} directory at HEASARC. GTI from the event file,
the background file, and the time intervals were merged, and spectra
were extracted from these times. Response files were then taken from
the \texttt{CALDB} database, and exposure times were adjusted to agree
with the spectra. The grouping of the \gso spectra is essentially
fixed by the grouping of the background file, thus no rebinning was
performed on these spectra. It was determined that the background
becomes prohibitive above 300\,keV; therefore, we restrict the \gso
spectra to the 60--300\,keV range.

\subsection{RXTE Analysis}\label{sec:data_rxte}

The \rxte data were prepared with tools from the HEASOFT v6.8 package
and the most current calibration files as of 2010 January 10. We used
standard filtering criteria for data from the Proportional Counter
Array \citep[\pca;][]{jahoda:06a}. Specifically, we excluded data from
within 30 minutes of South Atlantic Anomaly (SAA) passage, from
whenever the target elevation above the limb of the earth was less
than $10^\circ$, and from whenever the electron ratio (a measure of
the charged particle background) was greater than 0.15. We used the
background models appropriate to bright data.

We applied 0.5\% systematic errors to all \pca\ channels, added in
quadrature to the errors calculated from the data count rate. For all
\pca\ fits we grouped the data starting at $\ge 3$\,keV with the
criteria that the signal-to-noise (after background subtraction, but
excluding systematic errors) in each bin had to be $\ge 4.5$. We
restricted the noticed energy range to 3--22\,keV.

We extracted data from the High Energy X-ray Timing Experiment
\citep[\hexte;][]{rothschild:98a} using the same criteria as for the
\pca. \hexte is comprised of two clusters, $A$ and $B$. Prior to 2007,
both clusters were in a rocking mode, with only one detector viewing
the source at a given time, while the other detector conducted
off-source background measurements. For later dates (e.g., our three
most recent observations), \hexte A has been in a fixed on-source
position, which required us to estimate cluster A backgrounds using
the \texttt{hextebackest} tool.  Furthermore, during our third
observation there was clear evidence for a \hexte A exposure
anomaly\footnote{\texttt{http://gsfc.nasa.gov/docs/xte/whatsnew/}
  \texttt{newsarchive\_2007.html/}}.  We found evidence for such an
anomaly (albeit less severe) in our second observation as well. We
corrected the \hexte A exposure time by iteratively fitting an
exponentially cutoff powerlaw simultaneously to both cluster A and B
data and varying the exposure of cluster A until a $\chi^2$ minimum
was achieved using the same normalization for both detectors.

We binned the \hexte spectra to a common grid such that they had a
minimum combined signal-to-noise of 8 in each energy channel (even
though both clusters were fit individually). We further restricted the
noticed energy range to 18--200\,keV. Additionally, when fitting the
X-ray spectra we allowed the normalization of the \hexte backgrounds
to vary. (The best fit normalization constants were typically within
$\aproxlt 10\%$ of unity.)

For both \pca and \hexte spectra we further restricted the considered
time intervals to those that were strictly simultaneous with the
\suzaku spectra. That is, the \rxte time intervals form a subset of
the \suzaku time intervals.

\begin{figure}
\epsscale{1} \plotone{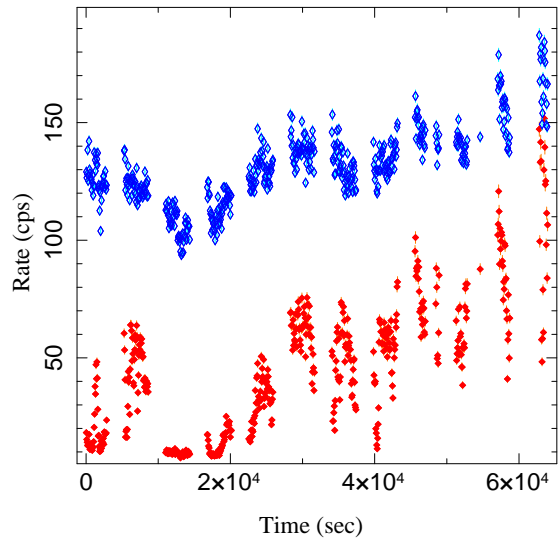}
\caption{\suzaku-\xis1 lightcurves (16\,sec integration bins) for
  observation 4 in the 0.5--1.5\,keV band (filled diamonds, bottom
  curve) and the 3--9\,keV band (hollow diamonds, top curve, shifted
  upward by 50\,cps).}
 \label{fig:lc}
\end{figure}

\subsection{Chandra-HETG Analysis}\label{sec:data_chandra}

The High Energy Transmission Grating \citep[\hetg;][]{canizares:05a}
was inserted for our observation of \cyg, with the data readout mode
being Timed Exposure-Graded. The \hetg\ is comprised of the {High
  Energy Gratings} (\heg), with coverage from $\approx 0.7$--8\,keV,
and the {Medium Energy Gratings} (\meg), with coverage from $\approx
0.4$--8\,keV. To minimize pileup in the gratings spectra, a 1/2
subarray was applied to the CCDs.  Additionally, the observatory
aimpoint was placed closer to the CCD readout. This configuration
reduces the frame time to 1.741\,sec, without any loss of the
dispersed spectrum.

\begin{figure*}
\epsscale{1} \plotone{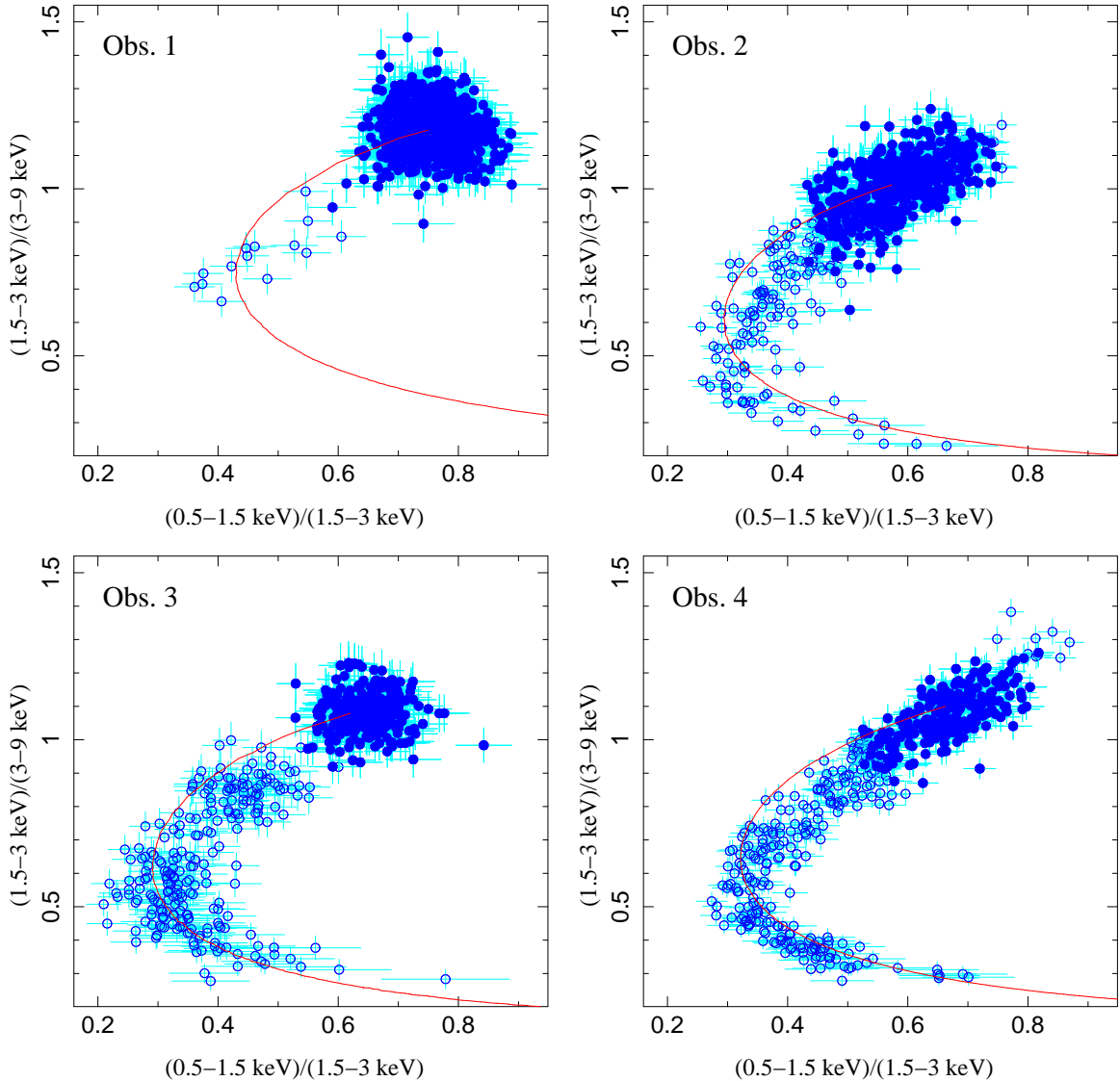}
\caption{\suzaku-\xis1 color-color diagrams, made with 16\,s bins.
  The colors are created from the count rate in three bands:
  0.5--1.5\,keV, 1.5--3\,keV, and 3--9\,keV.  The lines are the best
  fit dust halo models, presuming that the source is obscured by both
  an interstellar neutral column as well as by a neutral column local
  to the source, with the latter ranging in value from
  0--$2\times10^{23}\,{\rm cm^{-2}}$.  {Filled symbols indicate
    the periods used for the spectral analyses described in this
    work.} (See text.)}
 \label{fig:colors}
\end{figure*}

We used {\texttt{CIAO v4.2} and \texttt{CALDB v4.2.0} to extract
  the data and create the spectral response files.} The location of
the center of the \zth\ order image was determined using the
\texttt{findzo.sl}
routine\footnote{\texttt{http://space.mit.edu/ASC/analysis/findzo/}},
which provides $\approx 0.1$ pixel ($\approx 0.001$\,\AA, for \meg)
accuracy. The data were reprocessed with pixel randomization turned
off, but pha randomization left on. We applied the standard grade and
bad pixel file filters, but we did not destreak the data.

Although the instrumental set up was designed to minimize pileup, it
is still present in both the \meg\ and, to a lesser extent, the \heg\
spectra. We incorporate the effects of pileup in our spectral fits
using the \texttt{simple\_gpile2} model \citep{hanke:08a} based upon
the model originally described by \citet{nowak:08a}. For the small
amount of pileup present in these data, this model is extremely
successful in describing its effect on the spectra.

As for the \pca and \hexte spectra, we restricted the considered time
intervals for the \chandra spectra to those that were strictly
simultaneous with the \suzaku spectra.

\subsection{Estimated Radio fluxes}\label{sec:radio}

Simultaneous \ryle radio fluxes are not available for the X-ray
observations discussed in this work. Owing to the known stringent
correlations between the X-ray and radio flux, specifically the
correlations between the \rxte-\asm and \ryle fluxes (see
\citealt{gleissner:04b,nowak:05a}), we use an estimated radio flux
when applying the jet model to the spectra (\S\ref{sec:jet}).  From
our prior observations we average all radio data that occurred at
times with \asm daily averages that were within $\pm 1\sigma$ of the
\asm daily average corresponding to these new observations. The
average and standard deviation of these radio data were used in the
spectral fits, which yielded radio fluxes of $12.6\pm3.3$\,mJy
(2006-10-30), $10.8\pm3.9$\,mJy (2007-04-30), $7.9\pm3.0$\,mJy
(2007-05-17), and $10.2\pm3.0$\,mJy (2008-04-19).  Each average was
comprised of a minimum of ten measurements, and the extreme radio flux
values within each sample never differed from the average by more than
a factor of two.

\subsection{Data Plots}\label{sec:data_plots}

Throughout this work we will present spectra that are ``unfolded''
from the detector response in a model-independent manner
\citep[see][]{nowak:05a}. Specifically, we define the unfolded photon
flux, $F_{\rm unfold}(h)$, from a pulse height (PHA) bin, $h$, as
\begin{equation}
F_\mathrm{unfold}(h) = \frac{ C(h) - B(h) }{\int R(h,E)\,A(E)\,dE\,\Delta t} ~,
\end{equation}
where $C(h)-B(h)$ are the background subtracted counts, $R(h,E)$ and
$A(E)$ are the detector response matrix and effective area, and
$\Delta t$ is the exposure time.  For multiple data sets coadded
together for the plots (e.g., the \suzaku-\xis or the \rxte-\hexte
data), the numerator and denominator are each replaced with the sum of
the same quantities from the detectors.  Throughout we only sum data
for the plots, not for the fits.

Model spectra are ``unfolded'' in a similar manner. That is, predicted
background subtracted model counts are first calculated, and then
unfolded as above. Thus, the model spectra exhibit the smearing
induced by the detector response matrix. The one exception to this
model component scheme is for the jet results presented in
\S\ref{sec:jet}. There some of the model components are shown
unsmeared by the detector response, as they are derived from internal
calculations of the jet code.

\section{Lightcurves}\label{sec:lightcurves}

By design, our fourth observation occurred near binary orbital phase 0
(superior conjunction of the black hole), wherein we expected to be
viewing Cygnus X-1 through the wind of the secondary. By happenstance,
all of our remaining \suzaku observations also occurred near superior
conjunction, i.e., observations 1--4 cover orbital phases 0.2--0.3,
0.8--0.0, 0.85--0.0, and 0.0--0.14, respectively. Although dipping
events are known to be present during all orbital phases, during upper
conjunction dipping is prevalent due to absorption by clumps in the
wind \citep{balu:00a}. This behavior is clearly seen in the \suzaku
lightcurves presented in Fig.~\ref{fig:lc}. The soft X-ray lightcurve
shows a deep, prolonged $\approx 12$\,ksec dip.  A less pronounced,
but still significant dip is seen in the hard X-ray lightcurve
\citep{HankeMQW7}. Numerous dips of varying levels are also seen on
even shorter time scales. In order to extract the unabsorbed spectrum
of the source, it is therefore necessary to screen the data for
dipping and excise the dipping intervals from further analysis.

In addition to dipping, spectral modeling also has to take into
account the strong dust scattering halo in front of \cyg
\citep{predehl:95a}. As discussed by \citet[][and references
  therein]{xu:86a}, a scattering halo is produced by the scattering of
radiation from an X-ray source by a foreground dust cloud. Typical
halo sizes are on the order of arcminutes, i.e., comparable to the
size of the \suzaku PSF. The energy dependence of the scattering cross
section leads to a halo angular size that is $\propto E^{-1}$ and a
halo spectrum that is $\propto E^{-2}$.  Furthermore, as the scattered
photons travel along a greater path length to reach the observer, they
are delayed with respect to the directly observed photons by a factor
that is proportional to the square of the angular radius from which
the scattered photons are observed. The expected time delays for
nearby galactic sources typically range from thousands to tens of
thousands of seconds.

Thus the \suzaku spectrum and lightcurve of \cyg are comprised of two
components: a directly observed component, subject to the local
dipping events, and a time-delayed and time-averaged (due to
integration over different angular radii) softer spectrum from the
scattering halo. Most spectral analyses in the past have ignored the
effects of scattering since to first order for optically thin
scattering {in a homogeneous cloud}, whatever radiation that is
scattered out of the line of sight is scattered back in from larger
(often spatially unresolved) radii. Thus, for spatially unresolved (on
the size scale of the halo) and steady sources, dust can be ignored in
simple analyses.

If one assumes that the time-delayed and time-averaged spectrum is
represented by the same model as the direct spectrum, and if one
further assumes that interstellar absorption is predominantly in the
foreground of the dust halo, within \isis one can write the spectral
model {for the spectra during the dips} as:
\begin{eqnarray}
{\tt TBnew(1)*(1-dustscat*(1-TBnew(2)))*} 
   \nonumber \\{\tt (continuum~model)} ~.
\label{eq:dust}
\end{eqnarray}
Here and throughout we shall use an updated
version\footnote{\texttt{http://pulsar.sternwarte.uni-erlangen.de/wilms/}
  \texttt{research/tbabs}} of the absorption model of
\citet{wilms:00a} to describe both the interstellar (model instance 1)
and local (model instance 2) absorption. The \texttt{dustscat} model
is a version of the \xspec\ \texttt{dust} model that removes the
optically thin assumption of the latter model, but otherwise ignores
multiple scatterings (F. Baganoff, priv. comm.). Note that when local
absorption is absent ({\tt TBnew(2)} $\rightarrow 1$), the dust
scattering term drops out of the model expression.

Not all of the assumptions encompassed within the above model
expression are necessarily realistic. Specifically, the assumptions
that all interstellar absorption is foreground to the dust and that
the delayed source spectrum is identical to direct spectrum are
undoubtedly wrong in fine detail. However, we find that the above
expression provides a reasonable description of the color-color
diagrams in Fig.~\ref{fig:colors}. Specifically, we create a spectrum
from bright phases of the lightcurve that inhabit a locus of points in
the upper right hand corner of the color-color diagrams {(i.e.,
  the filled symbols of Fig.~\ref{fig:colors})}.  We fit a simple
spectral model to these selected data: an absorbed disk plus powerlaw
plus broad and narrow gaussian lines near 6.4\,keV. We then apply this
model in eq.~(\ref{eq:dust}), and create color-color curves by varying
the \texttt{TBnew(2)} component from
0--$2\times10^{23}\,\mathrm{cm}^{-2}$. These curves are then fit to
the color-color diagrams using the optical depth of the dust
scattering halo as the single free parameter. Our best fits yield
$\tau = 0.24$--0.34, and the results are shown in
Fig.~\ref{fig:colors}.

Overall, these curves describe the behavior of the color-color
diagrams reasonably well. The dust halo represents $\approx
20\%$--30\% (i.e., on the order of the scattering optical depth) of
the soft X-ray ($\aproxlt 2$\,keV) flux being ``uncovered'' during the
dips. Given the much narrower \chandra PSF, one would expect
comparable \chandra color-color diagrams to show a far smaller
uncovered fraction if the dust halo interpretation is correct. This is
indeed found to be true, with \chandra showing only an $\approx 2\%$
uncovered fraction \citep[][Hanke et al., in prep.]{HankeMQW7}.
However, it is possible that the very low uncovered fraction required
by the \chandra data might, in fact, be partly intrinsic to the local
dipping at the source (Hanke et al. in prep.). Note that the
theoretical curves show a greater degree of curvature than the
observed, nearly linear, evolution toward the lower left hand corner
of the diagram. We hypothesize that this discrepancy is due to our
assumption in eq.~(\ref{eq:dust}) of purely neutral absorption in the
dips. A concurrent increase in the optical depth of ionized absorption
could serve to alter the theoretical curves in the appropriate manner.
We return to this concept of an ionized absorber component in
\S\ref{sec:ion_line}.

\begin{deluxetable}{cccccc}  
\setlength{\tabcolsep}{0.03in} \tabletypesize{\footnotesize}
\tablewidth{0pt} \tablecaption{Observational Fluxes in keV Energy
  Bands \label{tab:flux}} \tablehead{\colhead{Date} & \colhead{0.5--2}
  & \colhead{2--10} & \colhead{10--100} & \colhead{100-300} &
  \colhead{${\rm L_{Bol}}$\tablenotemark{a}} \\ (yyyy-mm-dd) &
  \multicolumn{4}{c}{($10^{-8}~{\rm erg~cm^{-2}~s^{-1}}$)} & (${\rm
    L_{Edd}}$) } \startdata 2006-10-30 & 0.16 & 0.78 & 2.76 & 1.59 &
0.032 \\ 2007-04-30 & 0.07 & 0.55 & 2.07 & 1.32 & 0.026 \\ 2007-05-17
& 0.07 & 0.38 & 1.33 & 0.87 & 0.017 \\ 2008-04-19 & 0.14 & 0.72 & 2.62
& 1.44 & 0.035 \\ \enddata \tablecomments{Fluxes are absorbed values,
  are normalized to the \pca spectral fit, and correspond to the
  brightest/least absorbed periods of the lightcurves.}
\tablenotetext{a}{Unabsorbed, isotropic luminosity in 0.01-800\,keV
  band, expressed as a fraction of Eddington luminosity, as determined
  from the thermal Comptonization fits of \S\ref{sec:compton} and assuming a
  distance of 2.3\,kpc and a black hole mass of 10\,$M_\odot$.}
\end{deluxetable} 

\section{Simple Spectral Models}\label{sec:simple}

The lightcurves and color-color diagrams described above were used to
create spectra for the time periods corresponding to the bright phases
of the lightcurve represented in the upper right corner of the
color-color diagrams (Fig.~\ref{fig:colors}).  {The selections
  that we made are highlighted as filled symbols within these figures.
  The resulting exposure times are given in Table~\ref{tab:lo}.}  With
the exception of the \rxte-only broken powerlaw spectral fits
described immediately below, throughout the rest of this work we shall
only describe spectra from these bright phase, minimally locally
absorbed time periods.

\begin{deluxetable*}{ccccccccccc}  
\setlength{\tabcolsep}{0.03in} \tabletypesize{\footnotesize}
\tablewidth{0pt} \tablecaption{\texttt{constant*highecut*powerlaw}
 parameters for hard X-ray spectra fits
             \label{tab:cutoff}}
\tablehead{ Date
          & \colhead{$\Gamma$}
          & \colhead{$A_{\rm PL}$}                   
          & \colhead{$E_{\rm cut}$}
          & \colhead{$E_{\rm fold}$}
          & \colhead{$C_{\rm GSO}$}      
          & \colhead{$C_{\rm HEXTE~A}$}                        
          & \colhead{$C_{\rm HEXTE~B}$}      
          & \colhead{$B^{\rm norm}_{\rm HEXTE~A}$}      
          & \colhead{$B^{\rm norm}_{\rm HEXTE~B}$}      
          & \colhead{$\chi^2$/DoF}
          \\                               
          (yyyy-mm-dd) &&& (keV) & (keV) 
         }                                  
\startdata
         2006-10-30
         & \errtwo{1.41}{0.02}{0.01} 
         & \errtwo{1.07}{0.05}{0.03} 
         & \errtwo{19.6}{3.0}{2.7} 
         & \errtwo{183}{9}{5} 
         & \errtwo{1.007}{0.008}{0.008}
         & \errtwo{0.877}{0.004}{0.004} 
         & \errtwo{0.883}{0.005}{0.004}
         & \errtwo{1.12}{0.01}{0.02}
         & \errtwo{1.02}{0.02}{0.01}
         & 785.4/449 
\\ 
         2007-04-30
         & \errtwo{1.42}{0.01}{0.01} 
         & \errtwo{0.73}{0.02}{0.03} 
         & \errtwo{24.6}{1.8}{1.7} 
         & \errtwo{218}{7}{7} 
         & \errtwo{0.95}{0.01}{0.01}
         & \errtwo{1.07}{0.02}{0.03} 
         & \errtwo{1.07}{0.02}{0.01}
         & \errtwo{1.03}{0.10}{0.06}
         & \errtwo{0.99}{0.06}{0.06}
         & 966.2/315 
\\ 
         2007-05-17
         & \errtwo{1.47}{0.02}{0.02} 
         & \errtwo{0.60}{0.04}{0.03} 
         & \errtwo{23.5}{3.3}{2.7} 
         & \errtwo{252}{21}{18} 
         & \errtwo{0.95}{0.02}{0.02}
         & \nodata
         & \nodata
         & \nodata
         & \nodata
         & 301.6/123 
\\ 
         2008-04-19
         & \errtwo{1.39}{0.01}{0.01} 
         & \errtwo{0.93}{0.02}{0.03} 
         & \errtwo{20.4}{1.2}{1.3} 
         & \errtwo{164}{6}{5} 
         & \errtwo{0.91}{0.01}{0.01}
         & \errtwo{0.837}{0.003}{0.003} 
         & \errtwo{0.842}{0.004}{0.004}
         & \errtwo{1.065}{0.008}{0.008}
         & \errtwo{0.99}{0.01}{0.01}
         & 648.8/447 
\\
\enddata 
\tablecomments{Spectra are normalized to \hxd-\pin data. $E_\mathrm{cut}$,
 $E_\mathrm{fold}$ are the cutoff and folding energy of the
 \texttt{highecut} model. $A_\mathrm{pl}$ is the \texttt{powerlaw}
 normalization in $\rm photons\,keV^{-1}\,cm^{-2}\,s^{-1}$ at 1\,keV.
 $C$ are the fit constants for the detectors other than the \hxd-\pin. The
 normalizations of the \hexte backgrounds were also allowed to be
 free parameters, and are given by the $B^{\rm norm}$ values.}
\end{deluxetable*} 

\subsection{Broken and Cutoff Powerlaw Descriptions}\label{sec:cutoff}

As discussed by \citet{wilms:06a}, nearly all \rxte-\pca and \hexte
spectra of \cyg can be described by an \emph{extremely} simple
phenomenological model: an absorbed, exponentially cutoff, broken
powerlaw plus a broad, gaussian line. The photon indices of the broken
powerlaw, which we label as $\Gamma_1$ and $\Gamma_2$ for the soft
($\approx 3$--10\,keV) and hard ($\aproxgt 10$\,keV) X-rays, show a
correlation in the sense that the amplitude of the break between the
two, $\Gamma_1-\Gamma_2$, increases with higher (i.e., softer) values
of $\Gamma_1$.  This $\Gamma$--$\Delta \Gamma$ correlation is a very
phenomenological description of the spectral correlation that
elsewhere has been described as a ``$\Gamma$--$\Omega/2\pi$''
(hardness/reflection fraction) correlation
\citep{zdziarski:99a}. Following \citet{remillard:06a},
\citet{wilms:06a} classified spectra with $\Gamma_1\aproxlt 2.2$ as
``hard state'' spectra.

In order to compare the spectra discussed here to the work of
\citet{wilms:06a}, we take the 9 \rxte spectra\footnote{These 9
  spectra consist of three bright phase spectra, and six spectra taken
  from deeper dipping periods. Observation 1 shows no strong dips, and
  observation 3 does not have \rxte data strictly simultaneous with
  the bright phases of the \suzaku spectra.} and fit them with
exponentially cutoff broken powerlaws. As in the work of
\citet{wilms:06a}, these simple phenomenological models are excellent
descriptions of these data. This is even true for the spectra
extracted from times of moderate and deep dips. These latter spectra
have slightly reduced values of $\Gamma_1$ and increased values of the
fitted neutral column, $N_{\rm H}$ (\citealt{wilms:06a} show that
orbital phase-dependence of the column is discernible in our \rxte
observations of \cyg), but otherwise appear similar to non-dip phase
spectra. This is not entirely surprising given that \rxte is primarily
sensitive at energies $\aproxgt 3$\,keV, and therefore {does not
  measure} the spectral regime most strongly affected by the dips.

\begin{figure}
\epsscale{1}
\plotone{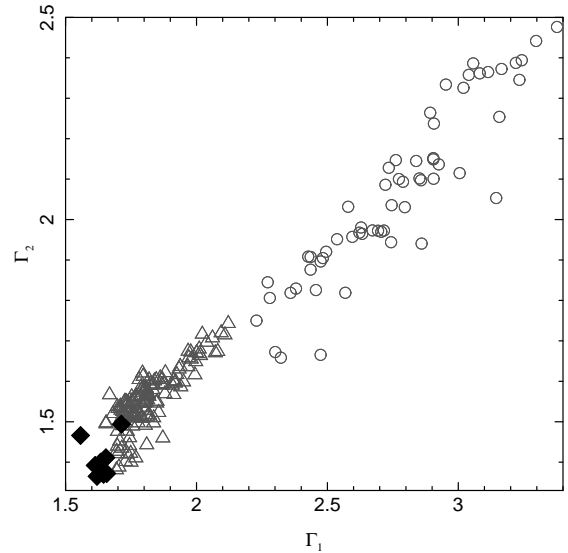}
\caption{Correlation of the soft X-ray ($\Gamma_1$) and hard X-ray
  ($\Gamma_2$) photon indices from exponentially cutoff, broken
  powerlaw fits to the \rxte \cyg data presented by
  \protect{\citet{wilms:06a}}.  Triangles represent the hard states
  ($\Gamma_1 < 2.2$), and circles represent soft states ($\Gamma_2 \ge
  2.2$). The solid diamonds represent exponentially cutoff, broken
  powerlaw fits to the \rxte data discussed in this work. There are
  spectra from four observations that have been further subdivided
  into up into times covering different portions of the color-color
  diagrams of Fig.~\protect{\ref{fig:colors}}, for a total of 9 \rxte
  spectra.}\label{fig:gamma_gamma}
\end{figure}

We show the resulting photon indices of these spectra along side the
results from \citet{wilms:06a} in Fig.~\ref{fig:gamma_gamma}.  Our
prior study covered seven years and 202 \rxte spectra.  These newer
observations overlap the historically hardest states measured in that
campaign. Specifically, for these new observations $\Gamma_1$ ranged
from 1.63--1.71, which is to be compared to the $\Gamma_1=1.65$
minimum found in \citet{wilms:06a}. In some cases, these observations
also correspond to historically faint hard states of \cyg. Defining
the ``hard state'' as $\Gamma_1 \le 2.2$, the lowest 2--100\,keV flux
in the hard state reported by \citet{wilms:06a} was
$2.46\times10^{-8}\,{\rm erg\,cm^{-2}\,s^{-1}}$, while the highest
2--100\,keV hard state flux reported in that work was
$4.50\times10^{-8}\,{\rm erg\,cm^{-2}\,s^{-1}}$. As shown in
Table~\ref{tab:flux}, the 2--100\,keV fluxes spanned by the bright
phase, least absorbed observations range from
1.71--3.54$\times10^{-8}\,{\rm erg\,cm^{-2}\,s^{-1}}$.

That these spectra are among the faintest and hardest ever observed
for \cyg is important to note.  In some models of the black hole hard
state, the hardest spectra are expected to have the smallest observed
reflection fraction and the disk inner radius should be at a maximum
\citep{zdziarski:99a}. However, it is also important to note that
other black hole systems can show substantially fainter spectra
relative to Eddington luminosity. Furthermore, as we see here and as
has been noted elsewhere, the range of hard state fluxes observed in
\cyg spans only a factor of two, and it is unlikely that the
bolometric flux spans a much greater factor than this even when
including observations from the soft state \citep{wilms:06a}.

\begin{figure}
\epsscale{1}
\plotone{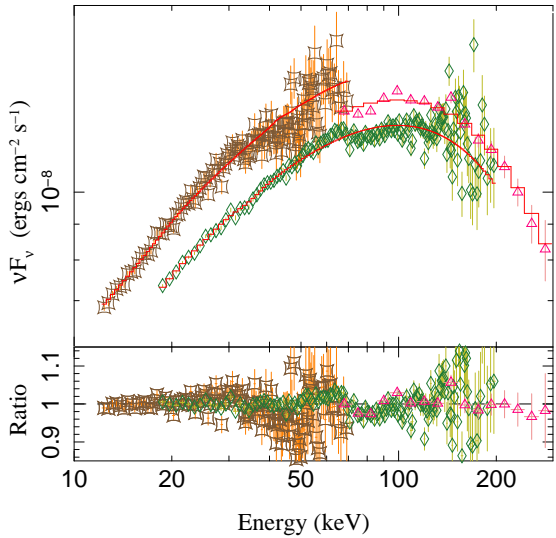}
\caption{\suzaku-\hxd \pin (brown squares) and \gso (magenta
  triangles) spectra, and \rxte-\hexte (green diamonds) spectra from
  the least absorbed periods of observation 4 fit with an
  exponentially cutoff powerlaw.}\label{fig:cutoff}
\end{figure}

As a further comparison to the previously observed hard states, we
consider the value of the fitted folding energy. For the hard states
discussed by \citet{wilms:06a}, the folding energy ranges from
$\approx 125$--255\,keV. Other BHC, e.g., GX~339$-$4, have shown a
wider range of hard state folding energies, i.e., $\approx
50$--300\,keV \citep{wilms:99a,nowak:02a,nowak:05a,motta:09a}. In our
prior studies we did not consider spectra above 125\,keV. Here,
however, with the inclusion of the \suzaku-\gso spectra we can now
consider energy ranges up to 300\,keV. For all four of our
observations, an exponential rollover is clearly detected. An example
is given in Fig.~\ref{fig:cutoff}, which also gives a general
indication of the consistency of the cross-correlation among the hard
X-ray detectors.  The values of the folding energies, along with
cross-normalization constants, are given in
Table~\ref{tab:cutoff}. The folding energies range from 164--252\,keV,
nearly the full span encompassed by our prior hard state
observations. This is not necessarily surprising. Whereas
\citet{wilms:06a} had observed a general trend for the cutoff to
increase to higher energies with harder spectra, there was a large
degree of scatter about this trend, consistent with these four
observations.

\subsection{Ionized Absorption Models}\label{sec:ion_line}

Previous \chandra-\hetg observations of \cyg have revealed evidence of
the ionized, focused wind from the donor star
\citep{miller:05a,hanke:08a}. The effects of this wind upon the
spectrum must be accounted for in any detailed analysis of \cyg. While
ionized line absorption is very pronounced during the deepest part of
the absorption dips (Fig.~\ref{fig:lc}), we stress that such
absorption is also significant during the bright phase (non-dip)
spectra discussed in this work.

\begin{figure}
\epsscale{1}
\plotone{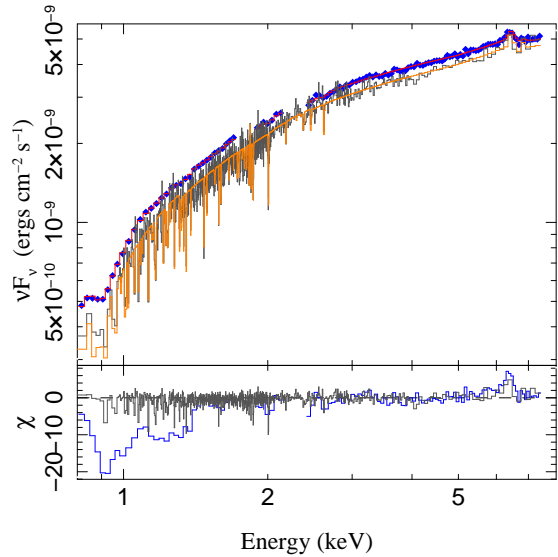}
\caption{\suzaku (blue diamonds) and \chandra-\hetg (grey histogram;
  $\pm$\fst order \heg only) data from observation 4. The model
  consists of absorbed disk and powerlaw components, broad and narrow
  gaussian lines in the Fe~K$\alpha$ region, and absorption lines from
  a variety of highly ionized species, including from Fe {\sc xxv} at
  $\approx 6.7$\,keV.  Additionally, we modify the \hetg spectra by
  dust scattering and a pileup term (see text).  The \suzaku and \hetg
  spectra have not been renormalized with respect to one another. The
  residuals shown here are for this joint model fit with all broad and
  narrow line normalizations (both emission and absorption) set to
  zero.}\label{fig:suzaku_chandra}
\end{figure}

We begin by considering the joint 0.8-7.5\,keV \suzaku and
\chandra-\hetg spectra from the 2008 April 19 observations. As the
\meg spectra are more heavily affected by photon pile-up, we fitted
only the \heg spectra. We used the same absorbed disk plus powerlaw
(\texttt{TBnew*(diskbb+powerlaw)}) continuum model to fit both the
\suzaku and \hetg spectra. We also added to both spectral models a
broad Fe K$\alpha$ line (modeled with \texttt{diskline};
\citealt{fabian:89a}) and a narrow Fe K$\alpha$ line modeled with a
gaussian, and a 6.7\,keV (Fe {\sc xxv}) absorption line. Following
\citet{hanke:08a}, we describe the line absorption as a series of
gaussian absorption lines representing a variety of ionized species ---
predominantly H- and He-like lines from elements ranging from O to Fe,
and L-shell transitions of Fe. This initial line list contained nearly
100 line transitions parameterized by wavelength (which we constrained
to fall within 1500\,${\rm km\,s^{-1}}$ of their rest wavelengths),
full width half maximum (constrained to be between 5--40\,mA), and
line equivalent width. Based upon fits to the joint \suzaku-\hetg
spectra, this initial list was reduced to the 55 significant lines
present in the spectra.

The above joint model received two modifications for solely the \hetg
spectra. The \texttt{simple\_gpile2} model (see \citealt{hanke:08a})
was applied to describe the effects of pile up in the gratings
spectra. Owing to this spectrum being fainter and harder than the \cyg
spectrum described in \citet{hanke:08a}, these spectra required an
even smaller pileup correction. Additionally, we applied the dust
scattering model (\texttt{dustscat}) to the \hetg spectrum. As
described in \S\ref{sec:lightcurves}, dust scattering represents a
loss term for the high spatial resolution \chandra spectra. This
component, albeit with a time delay of thousands to tens of thousands
seconds, scatters back into the \suzaku spectrum from arcsecond to
arcminute angular scales. To the extent that one can ignore the time
evolution of the spectrum, the dust scattering term can be ignored in
the \suzaku spectra. Lacking any detailed information of the average
spectrum prior to the start of our observations, we do not apply any
dust scattering correction to the \suzaku spectra.

{To be explicit, we apply to the \suzaku data a model of the form:}
\begin{eqnarray}
{\tt TBnew*lines*}
  \nonumber \\ {\tt (diskbb+powerlaw+diskline+gaussian) }
\end{eqnarray}
{while we apply to the \hetg data a model of the form:} 
\begin{eqnarray}
{\tt
  simple\_gpile2 \otimes
  (TBnew*lines*dustscat*}
   \nonumber \\ {\tt (diskbb+powerlaw+diskline+gaussian)) }~~,
\end{eqnarray}
{where \texttt{lines} represents the ionized line absorption.
  Throughout the rest of this work, we shall ignore as we are doing
  here the effects of dust scattering on the \rxte and \suzaku data.
  Again, this implicitly assumes that the bright phase spectrum has
  been steady over the time span of thousands to tens of thousands of
  seconds.  This assumption is unlikely to be valid in detail;
  however, we lack the data to employ any more sophisticated
  assumptions.}

This simple model describes the spectra well with a dust halo optical
depth of $0.25\pm0.03$ (90\% confidence level), i.e., consistent with
the fits to the color-color diagrams presented in
\S\ref{sec:lightcurves}. This spectral fit is shown in
Fig.~\ref{fig:suzaku_chandra}. The important point to note here is
that the ionized absorption is \emph{extremely} statistically
significant in the \suzaku spectra, even though the individual lines
are not resolved. Inclusion of these lines is vital for obtaining a
description of the soft end of the spectrum, and thus we include
ionized absorption in all of our model fits. Being that we do not have
\hetg spectra simultaneous with our other three \suzaku spectra, we
use the ionized absorption fit presented in
Fig.~\ref{fig:suzaku_chandra}. Specifically, for all subsequent
spectral fits described in this work, we freeze the line positions,
widths, and relative equivalent widths to the values found for this
joint fit. The line equivalent widths are tied together via a single
normalization constant, and this normalization constant becomes the
sole fit parameter describing the ionized absorption. The model,
however, still consists of 55 individual absorption lines. As
described below, this approach works well in all of our spectral fits.

\begin{figure}
\epsscale{1}
\plotone{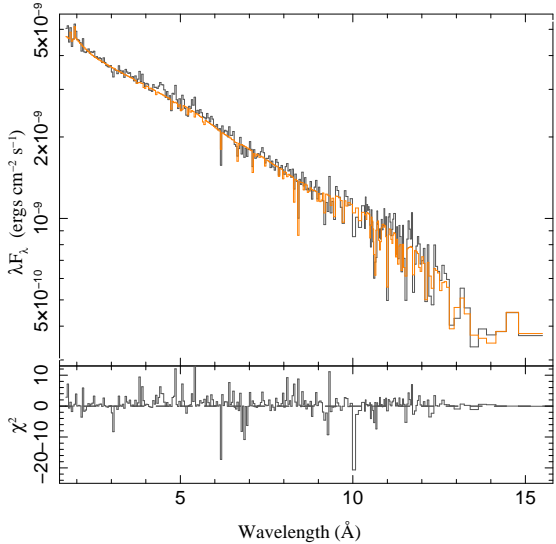}
\caption{{The \heg data from a joint fit of the \suzaku and \chandra
  data with a simple continuum model (disk, powerlaw, broad and narrow
  Fe region lines, neutral ISM absorption) plus two ionized absorber
  components described with the {\tt warmabs} model. The residuals
  include all of these models components; the remaining line
  residuals are features not described by the {\tt warmabs}
  components.}}\label{fig:warmabs}
\end{figure}

{We explored whether a more sophisticated ionization model could
  be applied to our data.  Rather than fit individual lines to the
  joint \suzaku and \chandra data, we instead used the {\tt warmabs}
  model \citep{kallman:01a} to describe the ionized absorption.  A
  model that includes two ionized components --- one with a column of
  $(7\pm4) \times 10^{21}\,{\rm cm}^{-2}$ and ionization parameter
  $\log_{10} \xi = 2.77\pm0.05$ and one with a column of $(3\pm2)
  \times 10^{21}\,{\rm cm}^{-2}$ and ionization parameter $\log_{10}
  \xi = 1.93\pm0.07$ (90\% confidence levels), and both with densities
  of $10^{10}\,{\rm cm}^{-3}$ --- describes the data reasonably well.
  As shown in Fig.~\ref{fig:warmabs}, however, a number of prominent
  line features are not well-fit in the \chandra data.  Furthermore,
  aside from being an extremely computationally expensive model to
  run, when fit to the \suzaku data alone the {\tt warmabs} models
  would tend to gravitate toward low ionization parameter values that,
  although capable of mimicking some of the continuum features of the
  \suzaku data, did not include the high ionization line features that
  were clearly present in the \chandra data.  For these reasons, we
  choose the empirical approach outlined above.}

{Several things need to be borne in mind when considering our
  results: whereas we allow for an overall normalization change in our
  ionized absorber model, we do not allow for changes of the
  ionization state.  As Fig.~\ref{fig:suzaku_chandra} shows, with the
  exception of a weak Fe {\sc xxv} line, there are essentially no
  ionized lines with energies $\aproxgt 3$\,keV.  Thus it is unlikely
  that this simple empirical approach will \emph{directly} impact, for
  example, broad Fe line studies.  However, in as much as ionization
  changes could affect our soft X-ray continuum fits, and these
  continuum fits in turn affect our estimate of the breadth and
  strength of the red wing of any broad line tail (see
  \S\ref{sec:rel_line}), potential ionization changes in this absorber
  must be considered an additional source of systematic uncertainty in
  the results discussed below.}

\subsection{Composite Relativistic Line}\label{sec:comp_line}

\begin{figure}
\epsscale{1}
\plotone{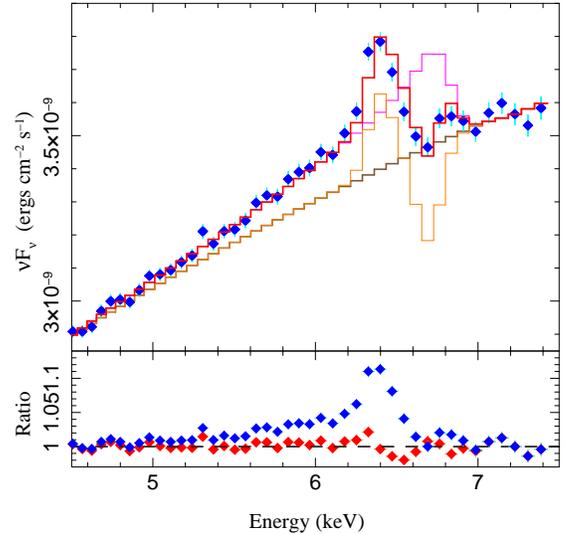}
\caption{The summed unfolded \suzaku\ spectra from unabsorbed periods
  from observations 2--4, shown in the Fe~K$\alpha$ line region.
  These data have been fit with a model consisting of an absorbed
  powerlaw, narrow Fe~K$\alpha$ emission line, narrow Fe {\sc xxv}
  absorption line, and a relativistically broadened \texttt{diskline}
  model. (See text for fitting procedures.)  The individual model
  components (powerlaw, with added diskline, and separately with added
  narrow emission and absorption lines) are shown overlain on the
  spectra, while the residuals are shown with and without the emission
  and absorption line components.}\label{fig:total_line}
\end{figure}

The fit described above consisted of both broad and narrow features in
the Fe K$\alpha$ region. Here we examine this spectral region in a
somewhat phenomenological manner to gauge the interplay between these
narrow and broad features. We wish to consider the average line
profile from roughly similar \cyg spectra. The three most recent
\suzaku observations occurred closest to orbital phase 0 and show very
similar color-color diagrams. As discussed below, these three spectra
are fit with comparable ionized absorption normalization constants. We
therefore consider these spectra jointly and ignore the first \suzaku
observation. We fit these spectra with a model consisting of a
powerlaw, narrow emission near 6.4\,keV and narrow absorption near 6.7
keV, plus a relativistically broadened Fe line (again using the
\texttt{diskline} model).  The relativistically broadened line is
characterized by a broad tail extending redward from the line rest
frame energy, and a sharper peak (due to Doppler boosting) blueward of
the line rest frame energy.

For purposes of this discussion, we allow the powerlaw to assume
independent parameter values for each of the \suzaku observations. The
energies and widths of the narrow lines are tied together for all
three observations, but the line strengths are fitted individually.
The broad line has an energy fixed to 6.4\,keV and an emissivity index
fixed to $\beta=-3$ (i.e., line emissivity is $\propto R^{-3}$, where
$R$ is the emission radius within the disk), and the disk inclination
is fixed at $35^\circ$ {\citep[the value adopted by][from the
    middle of the range of suggested inclinations]{herrero:95a}}. The
inner radius of the broad line emission is tied together for all three
observations, but the line strengths are left independent. The
$\chi^2$ of the fit is determined from the sum of $\chi^2$ values from
the individual observations; however, in Fig.~\ref{fig:total_line} we
show the spectra, fitted models, and residuals from the combined
data. We fit the spectra only in the 4.5--7.5\,keV bandpass.

\begin{figure}
\epsscale{1}
\plotone{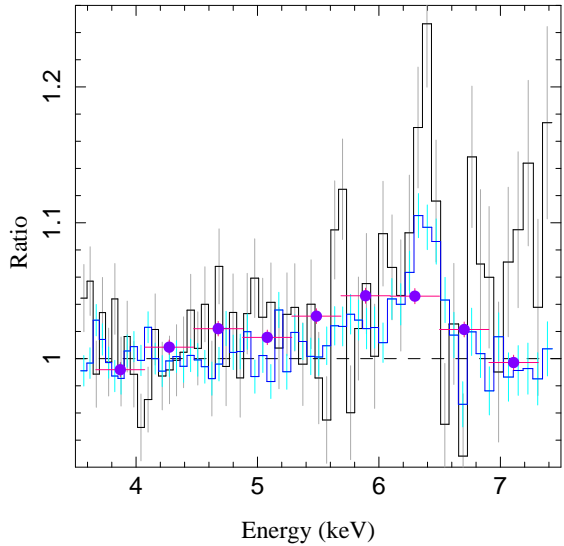}
\caption{\suzaku (blue line), \chandra-\heg (grey line), and
 \rxte-\pca (purple diamonds) residuals from fitting an absorbed
 powerlaw to the 3.5--4.5\,keV and 7--7.5\,keV regions
 simultaneously.  The \chandra-\heg data have been binned to match
 approximately the binning of the \suzaku data.
}\label{fig:line_residuals}
\end{figure}

These joint fits strongly require the presence of a broad line. The
broad line is significantly detected in each individual \xis detector
from each individual observation. Under the assumptions of the
continuum model applied here, the broad line equivalent width is
approximately three times that of the narrow line equivalent width.
The best fit inner radius for the broad line emission is
$10.6^{+2.9}_{-1.9}~GM/c^2$ (90\% confidence level), i.e., close to the
innermost stable circular orbit for a Schwarzschild black hole. As we
will discuss further in \S\ref{sec:compton}, the specific value for
such an inner radius is dependent upon the fitted
continuum. Nevertheless, a broad excess is required redward of the Fe
K$\alpha$ line.

In describing the residuals in this region with a broad line, we see
that the Fe {\sc xxv} absorption line occurs near the blue wing peak
of the relativistically broadened line (Fig.~\ref{fig:total_line}).
Both the narrow emission and absorption lines in the Fe region are
required in the \hetg spectra when considered by themselves (see also
\citealt{hanke:08a}), and thus they cannot be ignored in the \suzaku
spectra. The \hetg spectra, however, are not particularly well-suited
for describing any broad component of the Fe line. (See
Fig.~\ref{fig:line_residuals} and the discussion of
\citealt{hanke:08a} concerning broad line fits to our prior joint
\rxte-\chandra-\hetg observations of \cyg.)

\begin{figure}
\epsscale{1}
\plotone{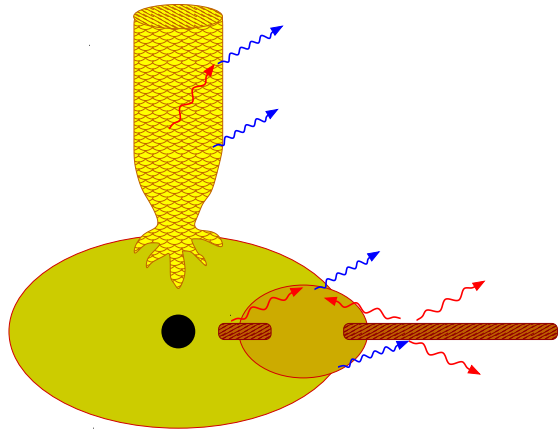}
\caption{{Possible X-ray emission geometry of the \cyg system.
    Part of the soft X-rays likely come from an accretion disk, which
    may be truncated at its inner edge by a hot corona.  Two possible
    (of many suggested) geometries are shown for this corona: a torus,
    and a quasi-spherical cloud.  In some models, the disk can reform
    in the inner regions of the accretion flow
    \protect{\citep{mayer:07a}}.  The corona will Compton upscatter a
    fraction of the disk component(s) soft X-rays to hard X-rays.
    Whether or not the corona envelopes the disk(s) determines whether
    or not an additional, unComptonized disk component should be
    present in the spectra.  (If the disk reforms on the inner edge of
    a toroidal corona, for example, a fraction of its soft X-rays will
    not intersect the corona.)  Hard X-rays will be reprocessed into
    reflection/fluorescent line components, predominantly by the outer
    disk. Finally, the jet might contribute to the observed X-ray
    emission via synchrotron and synchrotron self-Compton
    emission.}}\label{fig:geometry}
\end{figure}

On the other hand, if we compare the \rxte spectra to both the \suzaku
and \hetg spectra, we find that the \rxte spectra alone require an Fe
line equivalent width that cannot be accommodated with solely the
narrow line component from the \suzaku-\hetg fit. We highlight this
fact in Fig.~\ref{fig:line_residuals}, where we show the residuals
from a powerlaw fit in the 3.5--4.5\,keV and 7--7.5\,keV region to
the joint \rxte-\suzaku-\hetg spectra. In this figure, we have binned
the \hetg spectra to match approximately the binning of the \suzaku
spectra. There is overall good agreement between the \rxte and \suzaku
residuals; both require a broad red wing in the Fe line region. The
narrow Fe K$\alpha$ emission and Fe {\sc xxv} absorption are clearly
seen in the \hetg spectra; however, when viewed by itself, any broad
Fe line component in the \hetg spectra could be subsumed via a slight
powerlaw slope change. We ascribe this latter fact to remaining
calibration issues in the \hetg detector. Based upon these results, we
include broad and narrow Fe K$\alpha$ emission and narrow Fe {\sc xxv}
absorption in all the subsequent fits presented in this work. As we
shall discuss below, however, the implied broad line parameters are
dependent upon the assumed continuum model.

\section{Broad Band Models}\label{sec:bb}

We now turn to a discussion of the 0.8--300\,keV, joint \suzaku-\rxte
spectra for the bright phases of our four \cyg observations. Here we
consider three different Comptonization models, two of which are
discussed in detail, and a jet model. {(``Toy geometries'' for
  some of these situations are presented in Fig.~\ref{fig:geometry}.)}
The latter model also contains Comptonization components that
represent both Comptonization of disk photons as well as synchrotron
self-Compton (SSC). The non-jet models discussed below all rely upon
the \texttt{eqpair} model \citep{coppi:99a,coppi:04a} to describe the
fitted Comptonization components.

\begin{figure*}
\epsscale{1}
\plotone{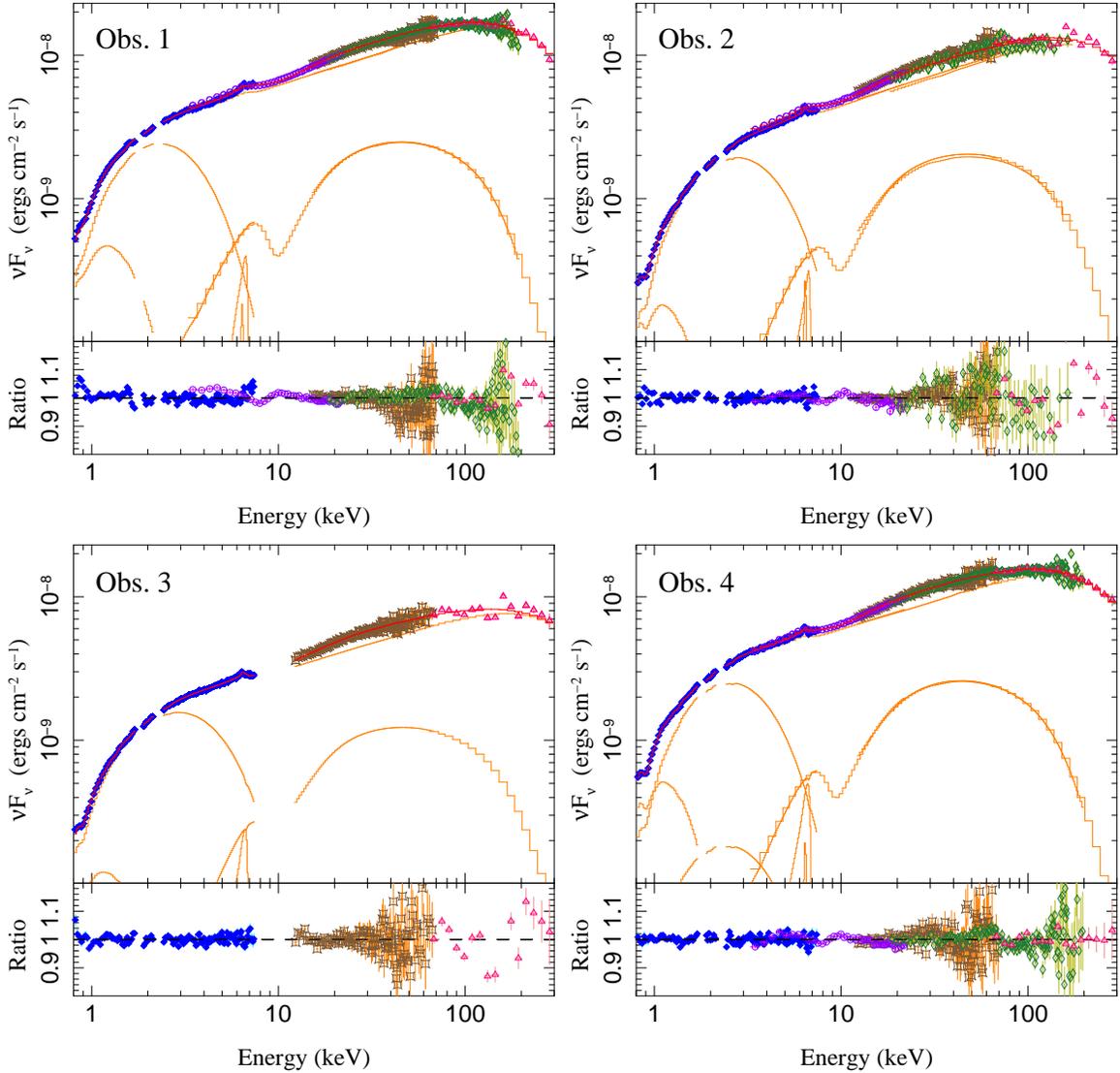}
\caption{Unfolded \suzaku and \rxte spectra, plus model components and
  fit residuals from the high $kT_\mathrm{seed}$, thermal corona
  \texttt{eqpair} fits. \suzaku-\xis spectra from individual
  detectors/data modes and \rxte-\hexte spectra from each cluster were
  summed for the figures, but not for the fits.  Here, and throughout
  the remainder of this paper, \xis data are filled blue diamonds,
  \pca data are hollow purple circles, \pin data are hollow green
  diamonds, \hexte data are brown squares, and \gso data are hollow
  magenta triangles. The following model components are shown: the
  summed model with the reflection component excluded, the reflection
  component by itself, the broad and narrow Fe K$\alpha$ line
  components, the seed photons without Comptonization applied, the
  additional disk component with peak temperature tied to the seed
  photons, and the additional low temperature disk component.  Neutral
  and ionized absorption are applied to all shown model
  components.}\label{fig:eqpair}
\end{figure*}

\subsection{Comptonization Descriptions}\label{sec:compton}

The \texttt{eqpair} model allows for Comptonization by a coronal
electron population that has both thermal and non-thermal energy
distributions. The latter distribution is governed by a parameterized
powerlaw, which for purposes of the fits described here we left at the
default \texttt{eqpair} values. Specifically, the electron phase space
density follows a distribution $\propto E^{-2}$, where $E$ is the
electron energy, between energies of 1.3\,eV--1\,MeV.  The eqpair
model rather than being parameterized by a coronal electron scattering
optical depth and temperature is instead parameterized by a
\emph{seed} electron optical depth (pair production within the corona,
accounted for in the code, can yield higher net scattering optical
depths) and coronal compactness parameters, $\ell$ (proportional to
component luminosity or power divided by radius). The latter
parameters are divided into two relative compactness parameters:
$\ell_\mathrm{h}/\ell_\mathrm{s}$, the coronal compactness divided by
the seed photon compactness, and $\ell_\mathrm{nt}/\ell_\mathrm{h}$,
the compactness of the non-thermal electron population divided by the
total electron population compactness. Over a wide range of
parameters, the hardness of the spectrum and the temperature of the
corona increases with increasing compactness (although pair production
can modify that effect).

\begin{figure*}
\epsscale{1}
\plotone{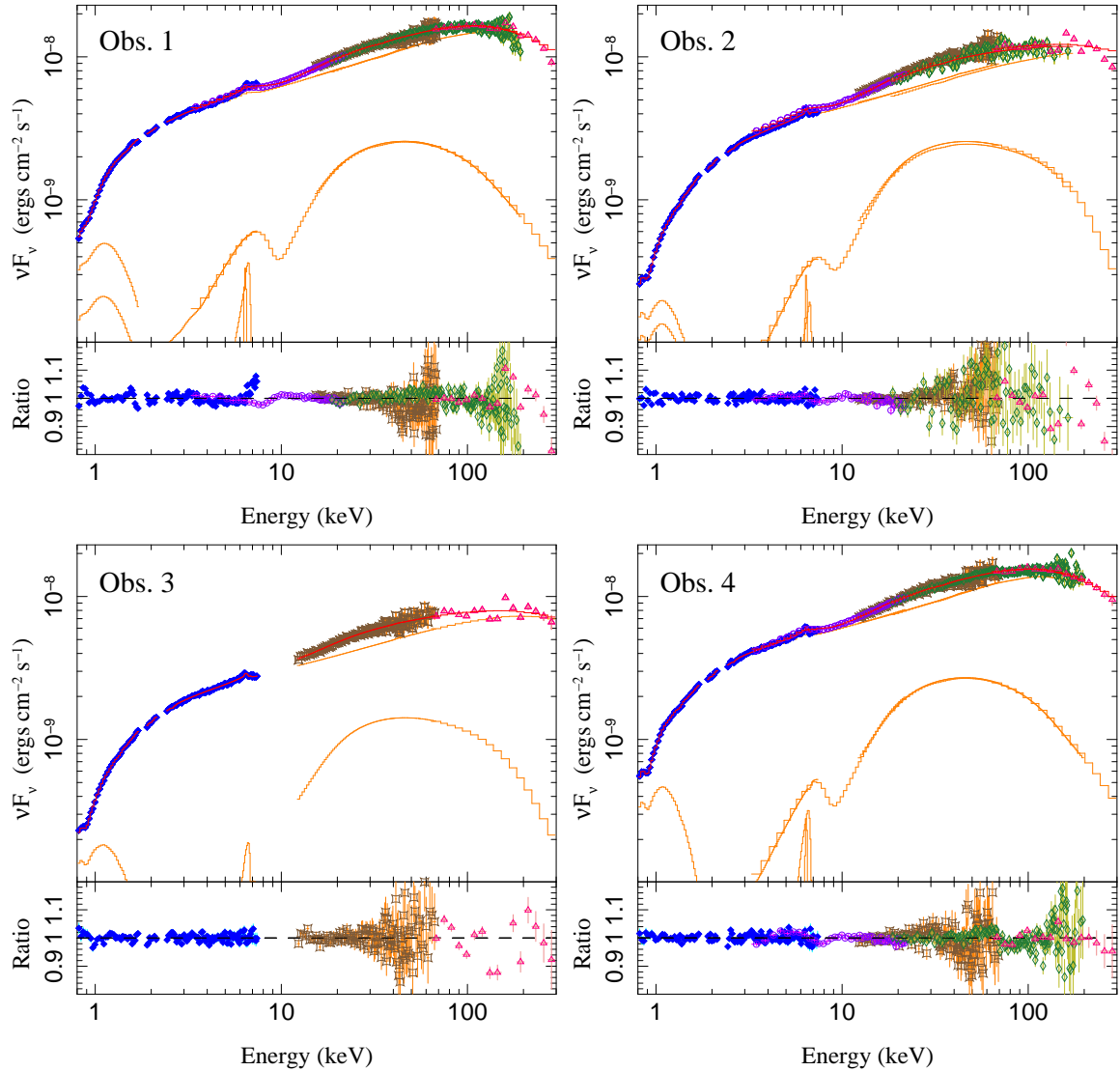}
\caption{Same as Fig.~\protect{\ref{fig:eqpair}}, but for the
  non-thermal corona, low $kT_\mathrm{seed}$ \texttt{eqpair} model
  fits. The following model components are shown: the summed model
  with the reflection component excluded, the reflection component by
  itself, the broad and narrow Fe K$\alpha$ line components, the seed
  photons without Comptonization applied, and the additional disk
  component with peak temperature tied to the seed photon peak
  temperature. Neutral and ionized absorption are applied to all shown
  model components.}\label{fig:nteqpair}
\end{figure*}

The seed photon spectrum can be set to either a blackbody or a disk
spectrum (essentially the \texttt{diskpn} model from \texttt{XSPEC};
\citealt{gierlinski:99a}). We choose the latter seed photon spectrum,
with the peak temperature of this disk ($kT_\mathrm{disk1}$) becoming
a fit parameter. We also allow for an unscattered disk component by
including an additional \texttt{diskbb} component with peak
temperature fixed to the peak temperature of the seed photons. Our
choice for the normalization of this component depends upon the
specific Comptonization model under consideration (see below).

To account for soft excesses (i.e., below $\approx 1.5$\,keV) in the
spectra, we sometimes include an additional unscattered disk component
with freely varying normalization and temperature ($A_\mathrm{disk2}$,
$kT_\mathrm{disk2}$).  We searched for, and easily found, solutions
where the peak temperature of this unscattered disk component was
below $\approx 300$\,eV.  The \texttt{eqpair} model includes
reflection from an ionized disk (i.e., a modification of the
\texttt{pexriv} model; \citealt{done:92a}) which is smeared by
relativistic distortions with the emissivity profile of a
Shakura-Sunyaev type disk \citep{shakura:73a}. A fittable parameter is
the inner radius of this disk smearing profile (see below). We also
fix the disk (i.e., reflector) temperature to $10^6$\,K, the reflector
inclination to $35^\circ$, and allow the reflection fraction
($\Omega/2\pi$) and reflector ionization parameter ($\xi$) to be fit
parameters. The latter is limited to values $\le 1000$.

Line absorption is described by the parameterized (i.e., single
normalization constant) model discussed in \S\ref{sec:ion_line}. A
narrow gaussian is added at 6.399\,keV with its width frozen to
0.01\,keV, but {with} a freely variable normalization. A
\texttt{diskline} component (essentially the profile for a disk around
a Schwarzschild black hole; \citealt{fabian:89a}) is added to describe
the broad line.  Its energy is fixed to 6.4\,keV, the disk line
emissivity index is set to $\beta=-3$, and the disk inclination is set
to $35^\circ$.  Typically one allows the inner emission radius to be
variable (see \S\ref{sec:comp_line}). Here we tie this inner radius to
that of the reflector in the \texttt{eqpair} model. The line
normalization, however, is allowed to freely vary.

A number of authors have considered ``sphere+disk'' Comptonization
models wherein an inner quasi-spherical corona is encircled by a
cool, geometrically thin disk that typically has a peak temperature of
$\approx 200$\,eV \citep{gierlinski:97a,dove:97b}. We were unable to
find any solutions for coronae with thermal electron distributions
(i.e., $\ell_\mathrm{nt}/\ell_\mathrm{h}$ frozen at $10^{-3}$) that
allowed for such a low seed photon temperature. The only such thermal
corona solutions that we found required seed photon temperatures of
0.8--1\,keV. Fit parameters for these solutions are presented in
Table~4, and spectral fits are shown in
Fig.~\ref{fig:eqpair}.

{Concepts for the potential geometry represented by this spectral
  fit are shown in Fig.~\ref{fig:geometry}.  The high temperature seed
  photons could be a recondensed inner disk as envisioned by
  \citet{mayer:07a}, while the low temperature soft excess, the broad
  line, and the reflection component could emanate from an outer
  disk. (The narrow line likely arises from fluorescence from the
  secondary and/or spatially extended gas surrounding the system;
  \citealt{torrejon:10a}.)  A more physically self-consistent model
  would allow for an additional seed photon contribution from this
  outer disk, and possibly a broad line component from the inner, hot
  seed photon emission region.  However, it is interesting to note
  that these} solutions required very little additional unscattered,
high temperature disk component.  {(Table~4
  contains low-- sometimes zero-- values of $A_\mathrm{disk1}$. Both
  the initial seed photon distribution and the associated high
  temperature, unscattered disk spectrum are shown in
  Fig.~\ref{fig:eqpair}; the former is much more significant.)  The
  amplitude of the soft seed photon contribution and the weakness of
  an additional unscattered high temperature disk would indicate a
  `recondensed' disk region of only modest extent, predominantly
  within the confines of a more extended corona, from which we would
  expect little iron line contribution.}

 On the other hand, these {spectral fits} do require an
 additional, unscattered low temperature disk component. The values of
 $A_\mathrm{disk2}$ presented in Table~4 correspond to
 inner disk radii of $\approx 2$--$10\,GM/c^2$, given a distance of
 2.3\,kpc, a black hole mass of 10\,$\msun$, and an inclination of
 $35^\circ$.  {This is roughly consistent with the inner radius of
   the fitted broad Fe line and relativistically smeared reflector,
   for which we find values ranging from $R_\mathrm{in}
   =6$--$18\,GM/c^2$. (The inner emission radius of the line appears
   to be uncorrelated with the inner radius implied by the
   normalization of the low temperature disk component.) These radii
   values are roughly} consistent with values close to the marginally
 stable orbit of a Schwarzschild black hole.

The broad line amplitude is between {6--22} times the amplitude of
the narrow line component. Fitted reflection fractions are
$\Omega/2\pi \approx 0.2$.

\begin{figure}
\epsscale{1}
\plotone{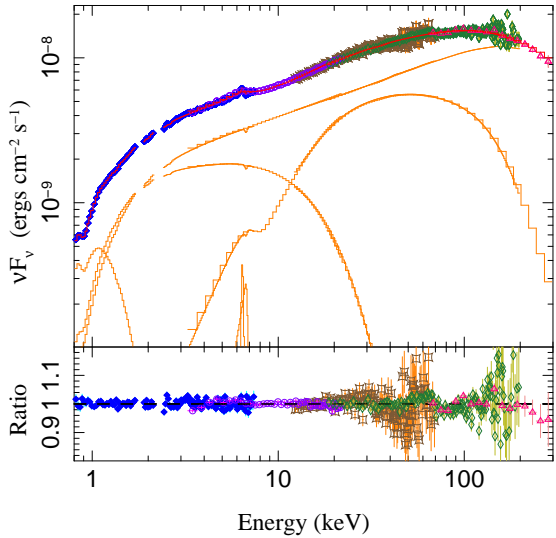}
\caption{Same as Fig.~\protect{\ref{fig:eqpair}}, but only showing
  observation 4 for the multiple \texttt{eqpair} component fit (see
  text).  The following model components are shown: the two individual
  eqpair components with their reflection components excluded, the
  reflection component of the dominant \texttt{eqpair} spectrum, the
  broad and narrow Fe K$\alpha$ line components, and the unscattered
  disk component.}\label{fig:deqpair}
\end{figure}

Comparable high seed photon temperature fits were found by
\citet{wilms:06a} when considering solely \rxte spectra. The major
difference between those fits and the ones discussed here is that our
prior fits also required a significant contribution from an
unscattered, high temperature disk. It is unclear whether that
difference is due to the inclusion of \suzaku data that extend to
lower energies, the inclusion here of ionized absorption, or the fact
that these spectra are for the most part fainter and harder than those
discussed by \citet{wilms:06a}.  The coronal compactnesses found here
are comparable to the maximum values from our previous studies, while
the seed optical depths are slightly lower (by $\approx 0.3$--0.5 for
comparable observations). These values of seed optical depth and
compactness correspond to coronal temperatures of 85--115\,keV and
total optical depths of 1.1-1.4.  The fitted reflection fractions are
slightly higher here (by $\approx 0.05$) than those fits with
comparable $\ell_{\rm h}/\ell_{\rm s}$ from \citet{wilms:06a}.

We have found a set of Comptonization model solutions that do allow
for a low seed photon temperature. For these fits, we include only one
extra unscattered disk component (with its peak temperature tied to
the peak seed photon temperature), and we further tie the inner disk
radius to the inner radii of the \texttt{eqpair} reflector and the
\texttt{diskline} emissivity. That is, we are explicitly mimicking the
``sphere+disk'' geometry (albeit not in a completely self-consistent
manner) {shown in Fig.~\ref{fig:geometry}}. If we then relax the
assumption of a purely thermal coronal electron distribution, good
descriptions of the spectra are found with low seed photon
temperatures. The ratio of non-thermal to total coronal compactness
then falls in the range $\ell_\mathrm{nt}/\ell_\mathrm{h} =
0.01$--0.82. That is, up to 82\% of the electron energy (presuming
that the thermal and non-thermal electrons are cospatial) resides in a
powerlaw distribution that extends to 1\,MeV.  Parameters for this
model are presented in Table~5 and spectra are shown
in Fig.~\ref{fig:nteqpair}.

The quality of these fits is similar to or perhaps slightly better
than that for the thermal corona fits. The fitted compactness
parameters for the non-thermal coronae are approximately double those
for the thermal fits. (Similar results were found by
\citealt{ibragimov:05a} when comparing thermal and non-thermal
\texttt{eqpair} fits.) At first glance, the optical depths seem
smaller; however, owing to the large coronal compactnesses pair
production is significant and the net optical depths range from
0.7--1.6, while the thermal electrons have temperatures that range
from 55--160\,keV.  Reflection fractions are slightly larger (by up to
0.09) for these fits. The broad line inner radius {in some cases}
has increased, and now ranges from $\approx
6$--32\,$GM/c^2$. Additionally, the amplitude of the broad line is now
only 3--13 times that of the narrow line. This difference compared to
the thermal corona model is due to two factors: the narrow line
amplitude is slightly increased, while the broad line amplitude in
{two} cases is nearly halved.

\begin{figure}
\epsscale{1}
\plotone{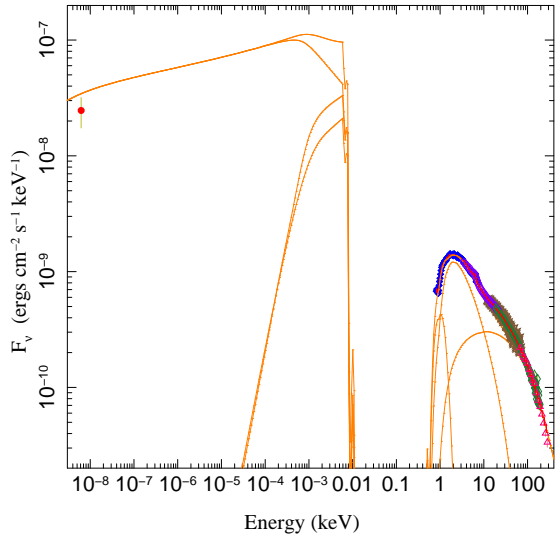}
\caption{The jet model fit to the estimated radio flux and measured
 X-ray spectra from observation 4.  Individual model components are
 shown with both neutral and ionized absorption applied. We show the
 summed model, as well as the the jet synchrotron component, the jet
 SSC component, the Compton corona component, and the disk component.
}\label{fig:alljet}
\end{figure}

We briefly consider one other thermal Comptonization solution that
allows for a low seed photon temperature. {(No ``toy geometry''
  for this model is shown in Fig.~\ref{fig:geometry}.)}  It is a model
similar to that considered by \citet{frontera:01a} when modeling
\beppo spectra of \cyg and to that considered by \citet{makishima:08a}
when modeling \suzaku spectra of \cyg.  In these models two {\tt
  eqpair} components are included.  They share the same seed photon
temperature (tied to the peak temperature of an additional unscattered
disk component, as before), and they share the same reflection
parameters; however, they have independent compactness and optical
depths.  An example of such a model fit is shown in
Fig.~\ref{fig:deqpair}.  These fits are generally successful, but no
more so than the previous two models.  Furthermore, owing to the
larger degree of freedom given the two semi-independent coronal
components, they are fraught with local minima representing
qualitatively different relative contributions of the two
Comptonization components.  We do not consider these more complex
models further in this work, other than to point out that such double
corona solutions as discussed by \citet{makishima:08a} and
\citet{frontera:01a} can allow for both low seed photon temperature
and purely thermal coronae.

\subsection{Jet Descriptions}\label{sec:jet}

The Compton corona models described above provide a good description
of the 0.8--300\,keV spectra.  These models, however, are not
unique. Even within the restricted class of coronal models, we have
shown three qualitatively different solutions that yield comparable
fits.  Furthermore, none of the Compton corona models give a
self-consistent description of the correlated radio spectra.  Although
the four observations discussed here did not have simultaneous radio
spectra, our prior studies \citep{gleissner:04b,nowak:05a,wilms:06a}
allow us to make good estimates of what the correlated 15\,GHz radio
flux likely was.  We describe this estimated joint radio/X-ray
spectrum with the jet model of \citet{markoff:05a} and
\citet{maitra:09a}.

\begin{figure*}
\epsscale{1}
\plotone{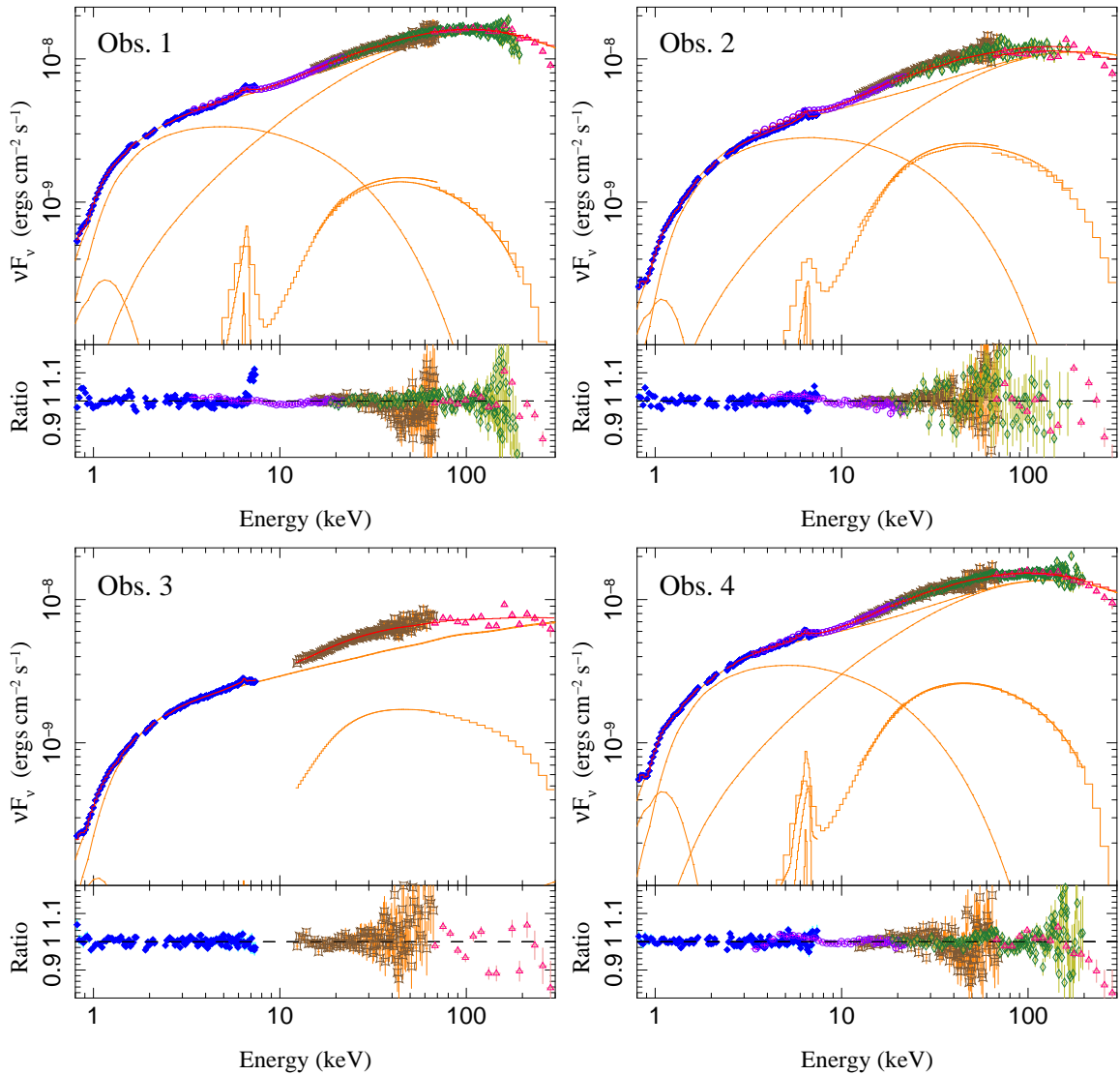}
\caption{Same as Fig.~\protect{\ref{fig:eqpair}}, but for the jet
  model fits.  The following model components are shown: the summed
  model with the reflection component excluded, the reflection
  component by itself, the broad and narrow Fe K$\alpha$ line
  components, the jet synchrotron component {(which dominates the
    2--10\,keV spectra for observations 1,2, and 4)}, the jet SSC
  component {(which dominates the 100--300\,keV spectra for
    observations 1,2, and 4)}, and the disk component. Neutral and
  ionized absorption are applied to all shown model
  components.}\label{fig:jet}
\end{figure*}

It is important to distinguish the jet models as discussed in Markoff
et al. (2004, 2005)\nocite{markoff:04a,markoff:05a} and later works
from the earlier iteration of the jet model discussed in Markoff et
al. (2001, 2003)\nocite{markoff:01a,markoff:03a}.  The 2001/2003 work
considered a parameter space where only synchrotron radiation from the
jet contributed to the observed X-rays.  {Later work}, and the
version of the jet model used here, include X-rays produced not only
by jet synchrotron radiation, but also by jet synchrotron self-Compton
processes and Comptonization of seed photons from the disk.  {The
  synchrotron component is only mildly beamed, with peak $\beta \sim
  2$ in the radio emitting portions of the jet.  The inverse
  Comptonization component comes from near the jet base where $\beta
  \sim 0.4$, and thus undergoes very little beaming.  Thus, although
  this does mean that the fitted jet parameters are dependent upon
  assumed inclination angle (e.g., the jet input power, $N_{\rm j}$
  will depend upon assumed inclination), given the low $\beta$ values
  inherent in the model it is possible to find good fit parameters for
  a wide variety of assumed inclinations.}

A full description of the main jet model parameters can be found in
the Appendix of \citet{markoff:05a} and in \citet{maitra:09a}.  The
version of the jet model that we use is closest to that discussed in
\citet{maitra:09a}; therefore, we repeat part of the model parameter
summary presented there.  We assume a distance of 2.5\,kpc and an
inclination of $35^\circ$ for the \cyg system. The jet properties are
then determined by the input jet power ($N_j$, expressed as a fraction
of the Eddington luminosity for a 10\,$M_\odot$ black hole), the
electron temperature of the relativistic thermal plasma entering at
the jet base ($T_e$), the ratio of magnetic to particle energy density
(the equipartition factor, $k$), the physical dimensions of the jet
base (assumed to be cylindrical with radius $R_0$ and height $h_0$,
the latter fixed to $1.5\,R_0$), and the location of the point on the
jet ($z_{\rm acc}$) beyond which a significant fraction of the leptons
are accelerated to a powerlaw energy distribution that follows
$E^{-p}$, with $p$ being a fit parameter.  The fit parameter
$\epsilon_{\rm acc}$ can be physically interpreted as being related to the
particle acceleration rate (which is $\propto \epsilon_{\rm acc}^{-1}$;
see \citealt{markoff:01a}).

As for the {corona-only} models, we also include emission from a
low temperature disk with inner radius $R_{\rm in}$ and peak
temperature $T_{\rm disk1}$.  {(This disk is very similar in
  temperature and normalization to the low temperature disk component
  included in the Comptonization model fits.)} A relativistically
broadened Fe K$\alpha$ line (with the same emissivity and inclination
parameters as for the {corona-only} models) is coupled to the disk
component by having its inner radius of emission tied to the disk
inner radius.  A narrow gaussian line is included, as well as a
reflection component (calculated with the {\tt reflect} model from
\texttt{XSPEC}).  Unlike the reflection model internal to the
\texttt{eqpair} code, the \texttt{reflect} model, which performs a
convolution on any given input spectrum, does not account for
relativistic smearing.  The Fe edge near 7.1\,keV in this unsmeared
reflected spectrum can produce a sharp feature not seen in the data
residuals; therefore, we smear this component with a unit-normalized,
$\sigma=1$\,keV gaussian convolution\footnote{Kernels for relativistic
  smearing do exist, e.g., the models of \citet{brenneman:06a} and
  \citet{dauser:10a}. The jet model calculations, however, are already
  substantially slower than the Compton corona calculations. Coupling
  them with the computationally expensive relativistic kernel would
  make the fits discussed here prohibitively time consuming to run.
  Furthermore, the jet model reflection geometry is likely more
  complex than that of the standard Compton corona: it should have
  separate contributions for the synchrotron, SSC, and Comptonized
  disk photons (see \citealt{markoff:04a}).  For these reasons we
  chose a very simple smearing profile.}.

Fit results for this model are presented in Table~6,
Fig.~\ref{fig:alljet}, and Fig.~\ref{fig:jet}.  The overall quality of
these fits is quite good, although not quite as good as for the
{corona-only} models discussed above.  The jet models leave slightly
larger residuals in the \pca spectra (near the Fe line region and the
10\,keV region), as well as in the \gso spectra.  The former could be
related to inadequate modeling of the reflection spectrum (i.e., not
using a relativistically smeared model).

Fits to the hardest X-ray spectra (i.e., the cutoff seen in the \gso)
are governed by the jet SSC and the inverse Compton components.  This
latter component is comprised of a magnetized, beamed corona with a
high temperature [$\approx (3$--$5)\times10^{10}$\,K, i.e., $\approx
3$--5\,MeV], and low electron scattering optical depth ($\tau_{\rm es}
\aproxlt 0.01$). A high temperature, low optical depth corona with a
limited residence time under those conditions simplifies the jet code
by allowing one to use a single scattering approximation for
calculating the Comptonization spectrum.  The compactness of such a
corona, however, could lead to high pair production which would then
serve to cool the corona and increase its optical depth (see, for
example, the critique of \citealt{malzac:09a}).

Post-facto estimates of the physical self-consistency of fitted jet
models are discussed by \citet{maitra:09a} who calculate the pair
production and annihilation rates at the base of the corona.  Fits to
the hard state spectra of XTE~J1181+105 yield far higher annihilation
rates than production rates, and therefore represent a self-consistent
solution.  Fits to the hard state spectra of GX~339$-$4 yield mixed
results.  For some cases, annihilation and production rates are
comparable and the resulting corona is marginally self-consistent.  In
other cases, the calculated production rates are an order of magnitude
higher than annihilation rates, and the resulting coronae are not
self-consistent \citep{maitra:09a}.

We have performed the same calculations as discussed by
\citet{maitra:09a} for jet models fitted in this work and find that
all yield pair production rates larger than pair annihilation rates.
The coronae within these jet models are therefore not self-consistent.
Observation 3 produces a fit that is {closest to being}
self-inconsistent, {with pair production and annihilation rates
  being a factor $\approx 10$ apart from one another. Jet} fits to the
other observations are further from self-consistency.

Such inconsistencies were similarly true for early coronal models,
e.g., those based upon the work of \cite{haardt:91a}.  A number of
such ``slab geometry'' coronae, in order to produce the hardest X-ray
spectra observed in sources such as Cyg~X-1, used coronal temperatures
(apart from any consideration of pair production) that were
unachievable in those geometries (see the discussion of this issue by
\citealt{dove:97a}).  Such inconsistencies were in fact what led to
the development of more physically self-consistent coronal models,
e.g., the \texttt{kotelp} model of \citet{stern:95a} and
\citet{dove:97b} and the \texttt{eqpair} model of \citet{coppi:99a}.

{We consider these jet models, however,} as they are as of yet the
only spectral models that make a serious attempt to explain the
correlated radio spectra.  Future iterations of these models will
incorporate consideration of pair production, as well as synchrotron
and SSC cooling of the Compton corona (Pe'er \& Markoff, in prep.).
(As shown by \citealt{dove:97a} and others, coronal cooling can be a
more significant consideration than pair production in some
situations.)

\begin{figure*}
\epsscale{1} \plotone{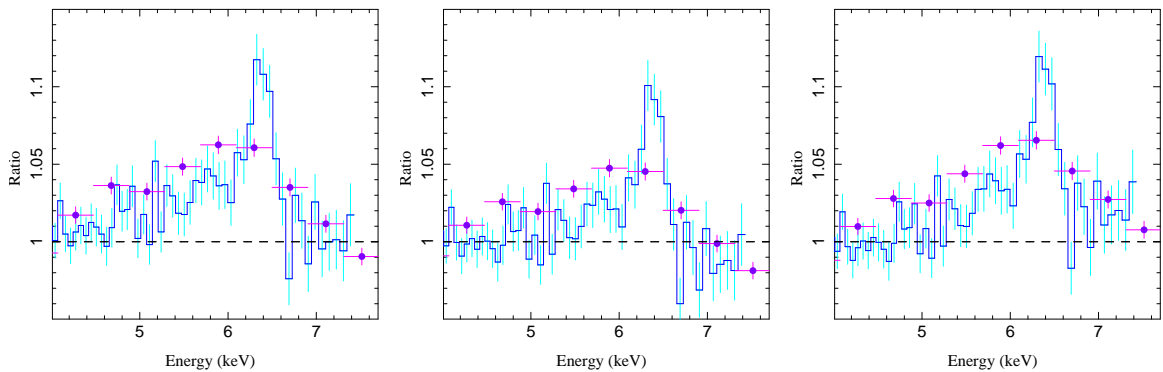}
\caption{Fe~K$\alpha$ line region residuals for observation 4,
  obtained from our best fits by setting the broad and narrow line
  normalizations (including the normalization of the Fe {\sc xxv}
  absorption line) to zero.  Left: the high $kT_\mathrm{seed}$
  \texttt{eqpair} fit, the hybrid thermal/non-thermal \texttt{eqpair}
  fit, and the jet model fit. (Purple circles are the \rxte residuals,
  while the blue histograms are the \suzaku
  residuals.)}\label{fig:line_comp}
\end{figure*}

The derived parameters are similar to those found for fits to other
BHC (see the fits to XTE~J1118+105 and GX~339$-$4 presented by
\citealt{maitra:09a}).  The power input at the base of the jet, in
terms of Eddington luminosity, is comparable to the inferred
bolometric luminosity (i.e., compare the values of $N_{\rm j}$ to the
values of Table~\ref{tab:flux}; {although different inclination
  assumptions will alter the value of $N_{\rm j}$, given the low
  $\beta$ values in the jet, it is always comparable to the inferred
  system luminosity}).  The radius of the base of the jet is
comparable to the radius of the innermost stable circular orbit
(ISCO).  The largest fit value exceeds the ISCO for a Schwarzschild
black hole by only a factor of 11.  The shock acceleration distance
along the jet, $z_{\rm acc}$ ranges from $\approx 60$--430\,$GM/c^2$.
The former value is comparable to that found for XTE~J1118+105, while
the latter is comparable to that found for GX~339$-$4.  As discussed
above, the fitted coronal electron temperatures at the base of the jet
range from $\approx (3$--$5)\times 10^{10}$\,K.  Work in progress
suggests that a physically self-consistent coronal temperature that
properly accounts for the dominant effects of synchrotron and SSC
cooling may truncate the allowed range of coronal temperatures to the
lower end found here (Pe'er \& Markoff, in prep.).  Future versions of
the jet code will explicitly account for these effects.

For all the fits presented in Fig.~\ref{fig:jet}, both synchrotron and
SSC radiation make substantial contributions to observed X-ray
spectra.  This is an important point to which we shall return in
\S\ref{sec:discuss}.

\subsection{Comparison of Implied Relativistic Lines}\label{sec:rel_line}

In Fig.~\ref{fig:line_comp} we show the Fe line region residuals for
the two Compton corona models and the jet model.  These residuals are
derived by setting the line normalizations (including absorption line
normalizations) to zero, \emph{without} refitting.  They are also
different from line residuals that are often used for illustrative
purposes where only a portion of the continuum outside of the line
region is fit with a simple powerlaw (e.g.,
Fig.~\ref{fig:total_line}).  The residuals in Fig.~\ref{fig:line_comp}
are meant to highlight the \emph{differences} in the derived line
parameters for the three main continuum models discussed here.

As shown by the parameters in Tables~4--6,
there are indeed differences among the inferred line parameters.  For
the most part, the purely thermal Comptonization model yields the
strongest line equivalent widths, while the non-thermal Comptonization
model yields the weakest equivalent widths (by {up to} a factor of
two).  The thermal Comptonization model {usually} also yields the
smallest inner radii for the relativistically broadened line, in some
cases extending all the way to the ISCO {for a Schwarzschild black
  hole}.  Larger values of these radii, however, are typically found
with the other two models.  These differences are evident in the lower
energy residuals shown in Fig.~\ref{fig:total_line}.

To be clear, a substantially broadened line is \emph{required} by
\emph{all} spectral models that we have explored, and there is a good
agreement between the \pca and \suzaku-\xis residuals regarding this
fact.  {We also reiterate that for all explored models, the
  equivalent width of the fitted narrow line (typically $<15$\,eV) is
  only a fraction of that for the broad line, and is in fact
  substantially smaller than the Fe line equivalent width we have
  found for any model that we have applied to any \rxte observation of
  \cyg \citep{wilms:06a}.}  Furthermore, {for these four
  observations}, we have not found any implied inner radius greater
than $\approx 40\,GM/c^2$.  The precise details of this broadened
line, however, do depend upon the assumed continuum model.

\section{Discussion}\label{sec:discuss}

In this work we present broad band (0.8--300\,keV) fits to four
separate observations of \cyg that have simultaneous \suzaku and \rxte
data. The most recent of these observations also has simultaneous
\chandra-\hetg data.  Each of these observations shows evidence of
dipping events likely due to dense structures (``clumps'') in the
otherwise highly ionized wind of the secondary.  This is seen in the
lightcurves (Figs.~\ref{fig:lc}), and the color-color diagrams
(Fig.~\ref{fig:colors}, which {can be modeled by} the presence
of a dust scattering halo).  The ionized absorption is very
significant in the \suzaku spectra (Fig.~\ref{fig:suzaku_chandra})
even though in this work we consider the least absorbed {periods}
of the lightcurve.
\begin{itemize}
\item Modeling the ionized line absorption present in \cyg, via the
 use of parameterized fits to the \hetg spectra, is crucial for
 deriving good fits to the soft X-ray spectra seen by \suzaku and
 \rxte.
\end{itemize}

We note that even accounting for this ionized absorption, the spectral
fits presented in this paper yield reduced $\chi^2_\nu$ that range
from 1.8--2.5.  Do such values truly represent good fits to these
data?  \cyg is bright, and these observations are of sufficient length
that the signal-to-noise values for these spectra are quite high.  The
spectra are dominated by systematic errors, especially at the soft
X-ray energies.  We already have added 0.5\% systematic errors to the
\pca spectra, which is a reasonable estimate for the \emph{internal}
uncertainty of the \pca.  There are also relative uncertainties among
the detectors. For observation 4, if we increase the \pca systematic
errors to 1\%, add 1\% systematic errors to the \hexte A cluster
(i.e., the fixed cluster), and add 3\% systematic errors to the
\suzaku-\xis spectra, then the fits presented here would have reduced
$\chi^2_\nu \approx 1$.  Comparing the fits among the individual \xis
spectra (representing both different individual detectors and
different data acquisition modes), we have found that $\pm 3\%$ is a
reasonable estimate of the end-to-end differences among these spectra.
We hypothesize that the quality of the fits presented here is near
``optimal'' given the current internal and relative calibrations of
these detectors.

{The observations that occurred on 2007 May 17, i.e., the third
  set of observations, are potentially the most problematic in turns
  of cross-calibration issues.  For our other sets of observations,
  the \rxte-\pca data act as a ``bridge'' between the \suzaku-\xis and
  -\pin spectra. Each can have its relative normalization anchored by
  a comparison to the \pca, which overlaps the energy coverage of both
  detectors.  This is lacking in the third observation, leading to the
  worry that changes in the normalization constant are subsuming, for
  example, continuum spectra associated with the spectral break at
  $\approx 10$\,keV.  For the third observation, we find the ratio of
  the \pin to \xis1 normalization constant to range from 1.10--1.17.
  This is to be compared to the 1.16--1.18 value found for fits to the
  first observation, the 1.06--1.08 found for the second observation,
  and the 1.10--1.11 found for the fourth observation.  The expected
  value\footnote{{\tt
      http://heasarc.gsfc.nasa.gov/docs/suzaku/analysis/} {\tt
      watchout.html.}  Since we did not fit \xis0 data for the fourth
    observation, here we compare to \xis1.} for the \pin/\xis0
  comparison is 1.16.  There is some amount of scatter for the
  cross-normalization values among the different observations;
  however, the first, second, and fourth observations show little
  scatter in its value for different fits to the same data.  More
  scatter is seen for the third observation; therefore, additional
  systematic uncertainties need to be considered as being present for
  that set of observations.}

Using simple broken powerlaw and exponentially cutoff powerlaw fits,
we find that these spectra are among the hardest seen in the
``low hard state'' of \cyg over the past decade.  For all four
observations, the spectra are clearly detected out to 300\,keV with
the \gso, and exponential cutoffs are well constrained.
\begin{itemize}
\item Although these are among the hardest \cyg spectra ever detected,
 the exponential folding energies vary by over a factor 1.5, and range
 from 160--250\,keV.  
\end{itemize}
Historically observed hard state spectra in \cyg have shown folding
energies that vary over a slightly wider range, while hard state BHC
as a class show folding energies that span a factor of 5.

As these spectra are among the faintest and hardest for \cyg, they make
excellent test beds for theoretical models that posit, for instance,
that the hard state represents a configuration with an inner disk that
has evaporated into a quasi-spherical corona.  In such a scenario, we
might expect these spectra to show the most ``extreme'' recession of
the inner disk, although we have noted that the bolometric
luminosities represented by these spectra only span a factor of two.
At a few percent of the Eddington luminosity, they are not far below
the expected soft-to-hard state transition.  Numerous transient BHC
sources show much fainter hard states as they fade into quiescence.

We have presented a number of different spectral models, all of which
describe the 0.8--300\,keV spectra well.  Some of these models
describe the X-ray spectra primarily with Comptonization components
(whether due to a thermal or hybrid thermal/non-thermal corona), while
the jet model is dominated by synchrotron and SSC emission from the
jet.  All of these models have a number of features in common.
\begin{itemize}
\item All models require a soft excess that here we describe
 with a disk component with low ($kT_{\rm disk} \approx 200$\,eV)
 peak temperature.  The implied inner radii of these disks 
 range from 2--40\,$GM/c^2$.
\item All models require a relativistically broadened line component.
 The inner emission radius of this broadened line never exceeds
 $\approx 40~GM/c^2$, but for some models is as low as 6\,$GM/c^2$.
\item All models require a reflection component.  The typical values
 for the reflection fraction are $\Omega/2\pi \approx 0.2$--0.3.
\item All models imply that the spectral hardening at $\approx
  10$\,keV is not \emph{solely} due to reflection.
\end{itemize}

This latter point is very important, and broadly agrees with similar
conclusions drawn by \citet{frontera:01a}, \citet{ibragimov:05a}, and
\citet{makishima:08a}. The presence of the broad Fe line and
\emph{some} of the spectral curvature in the 20--300\,keV band is a
clear indication of the presence of reflection.  However, this
reflection spectrum is \emph{not} sitting on top of a simple
``\texttt{disk+powerlaw}'' spectrum.  There is additional continuum complexity
separate from reflection that contributes to this perceived break.  A
high seed photon temperature in the thermal corona model yields a soft
excess in the 2-10\,keV band and thus contributes to the measured
break in that scenario.  (See the discussion in \citealt{wilms:06a}.)  As has been
discussed by \citet{ibragimov:05a}, a non-thermal electron population
in the corona can lead to a soft excess in the
2--10\,keV band that helps contribute to what otherwise would be
modeled as a reflection break at 10\,keV.  The two corona model
discussed by \citet{makishima:08a}, a version of which is shown in
Fig.~\ref{fig:deqpair}, rather explicitly replaces part of the
reflection component with a broad band continuum model.  Finally, in
the jet paradigm the spectral break at 10\,keV is partly
attributable to the transition from dominance by synchrotron emission
to SSC emission in the continuum.

There are plausible physical scenarios for each of the discussed
spectral models, {with some hypothesized geometries being shown in
Fig.~\ref{fig:geometry}}.  The hybrid thermal/non-thermal coronal model
is the closest to the concept of the quasi-spherical inner corona with
outer geometrically thin disk
\citep[e.g.,][etc.]{eardley:75a,shapiro:76a,ichimaru:77a,dove:97b}.
We have only been able to find such solutions, however, when invoking
a partly non-thermal electron population in the corona.

The purely thermal coronal model contains two disk components that are
reminiscent of the physical description given by \citet{mayer:07a}.
These authors describe a situation where an outer, geometrically thin,
cool disk surrounds an inner, geometrically thick, hot corona.  In the
very inner radii of this corona, however, thermal conduction leads to
it condensing into a geometrically thin and optically thick disk {(see
  Fig.~\ref{fig:geometry})}.  Such a component could supply the high
temperature seed photons in our thermal corona solutions, {while the
  lower temperature outer disk could provide the bulk of the
  reflection features}.  The jet model has a natural physical
interpretation in that the usually observed optically thick radio
spectrum observed in the hard state is clear indication of the
presence of a jet.  The question that remains is the contribution of
this component to the X-ray band.

The fact that the continuum is more complex than a simple
``\texttt{disk+powerlaw}'', yet there are multiple, physically
motivated models that yield comparably good spectral fits, leads to
the final point.
\begin{itemize}
\item Although a relativistically broadened line is required in all of our 
 spectral models, the parameters of this line are dependent upon the
 presumed continuum model.
\end{itemize}
{Coupled with this dependence upon assumed continuum spectrum is
  an implicit dependence upon ionized absorption, for which we have
  detailed \chandra-\hetg measurements for only the fourth
  observation\footnote{Since performing these observations, we have
    carried out a \chandra-\hetg observation of orbital phase
    0.5 (PI: Nowak), and have an approved observation of orbital
    phase 0.25 (PI: Hanke).  Coupled with archival \chandra-\hetg
    observations of orbital phases near 0.75, we hope to develop a
    better understanding of how the ionized absorption evolves with
    orbital phase, which should allow us to improve our spectral
    modeling in the future.}.}  Again, we have not found an inner
radius for this line that exceeds $\approx 40~GM/c^2$. Given the
variations of this line with presumed continuum, however, we are as of
yet unable to use this line for more refined diagnostics such as
estimates of black hole spin.

Although we are unable as of yet to draw firm conclusions as to the
best geometry and physical mechanisms to describe the hard state
spectra of \cyg, these new joint \suzaku-\rxte data provide a stunning
contrast to our prior results using solely \rxte data
\citep{wilms:06a}.  For the \rxte data alone we were able to describe
the 3--125\,keV spectra with a variety of physically motivated
Comptonization models and to describe correlations among the fit
parameters.  On the other hand, the simple exponentially cutoff,
broken powerlaw models with a single, broad gaussian line described
the data equally well, if not better \citep{wilms:06a}.  When
considering the 0.8--300\,keV \suzaku-\rxte data discussed here, this
is no longer the case.  We now require complex absorption at low
energy, an asymmetric broad line plus narrow emission and absorption
components in the Fe line region, and a complex continuum model.  The
catalog of \suzaku observations of BHC will continue to increase such
that we observe a wider variety of BHC states and luminosities.
Furthermore the sophistication and physical self-consistency of the
spectral models will continue to improve.  Together, they offer the
promise of obtaining a better understanding of the physical processes
occurring in these BHC systems.

\acknowledgements We would like to thank the \rxte, \suzaku, and
\chandra schedulers for making these simultaneous observations
possible.  This work was supported by \suzaku and \chandra guest
observer grants, NNX07AF71G, NNX08AE23G, NNX08AZ66G, GO8-9036X, as
well as NASA Grant SV3-73016.  M. Hanke and J. Wilms would like to
acknowledge the support of the BMWi through DLR grant 50 OR 0701.  The
research in this work has been partially funded by the European
Commission under grant ITN 215212.  S. Markoff and D. Maitra
acknowledge support from a Netherlands Organization for Scientific
Research (NWO) Vidi and OC Fellowship, respectively.

%\bibliographystyle{jwapjbib}
%\bibliography{mnemonic,jw_abbrv,apj_abbrv,bhc,agn,diplom,inst,ns,conferences}

\appendix

\section{\suzaku Attitude Correction and Pile-up Estimation}\label{sec:appendix}

We have created two tools to aid our \suzaku data analysis: {\tt
 aeattcor.sl} and \texttt{pile\_estimate.sl}.  The tools, descriptions
of their use, and example results can be found at: {\tt
 http://space.mit.edu/ASC/software/suzaku}.  The former tool further
corrects the \suzaku attitude solution, whereas the latter estimates
the degree of pileup in a given subset of the observation.  Both are
scripts written in \texttt{S-Lang} but are designed to be run on a
terminal command line using \texttt{ISIS}, the Interactive Spectral
Analysis System \citep{houck:00a}, as a driver.  For all intents and
purposes, the scripts behave similar to the typical Unix command line
tools found in the \texttt{HEASOFT} package.

\begin{figure*}
\epsscale{0.99} \plotone{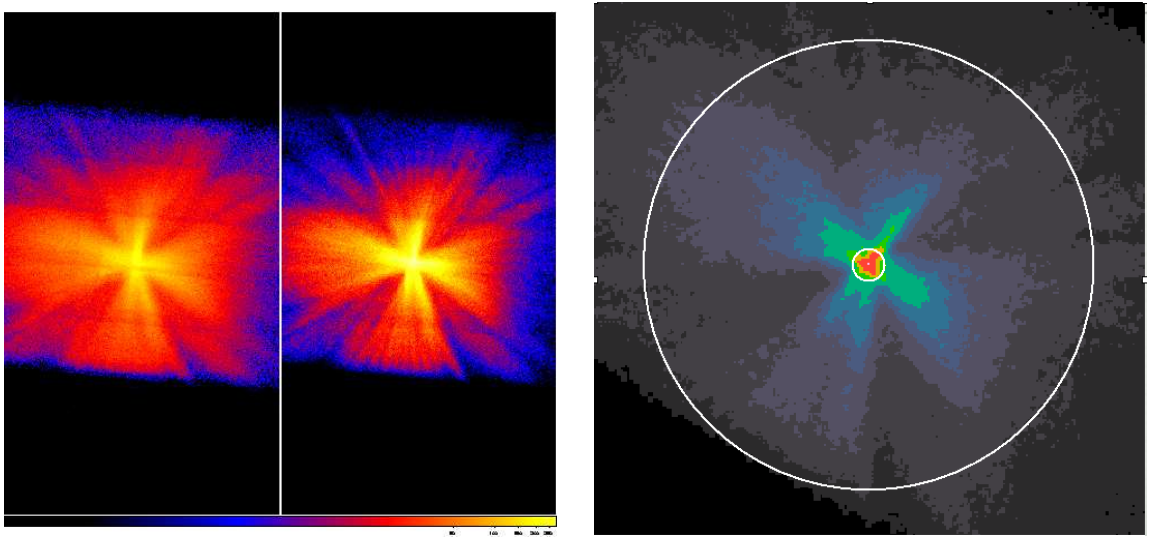}
\caption{Left: An example of \suzaku attitude correction. The left
  half of the image shows a \cyg image using the standard attitude
  correction. The right half of the image shows the improvement with
  \texttt{aeattcorr.sl}.  Right: An image of \cyg where discrete
  colors correspond to the pileup in that region.  The outer white
  circle denotes the outer boundary of an annular extraction region,
  and the inner white circle denotes the inner boundary. Pileup
  fractions within this excluded region are as high as 35\%. The
  average effective residual pileup level is
  $<4\%$.}\label{fig:correction}
\end{figure*}

Thermal flexing of the \suzaku spacecraft leads to a slow wobbling of
the optical axis, and hence blurring of the image. Current \suzaku
tools partially correct this effect by adjusting the spacecraft
attitude file based upon details of the spacecraft orbit, temperature,
etc. \citep{uchiyama:08a}. \texttt{aeattcor.sl} further improves this
correction (Fig.~\ref{fig:correction}) for a bright source by using
its time-dependent detected image to create a new attitude file.
Specifically, the tool attempts to shift the time-dependent mean
detector image position to a new specified, fixed sky position. It
presumes that there are no intrinsic variations in the time-dependent
image position. (A mean image position may be time varying, for
example, if it is comprised of two or more variable sources of
comparable flux.  Additionally, a highly variable and piled up source
where an ``image crater'' appears and disappears over the course of
the observation may lead to a variable mean image position.)  The bin
time over which the tool searches for and shifts the image peak is a
user selectable parameter; however, we have used the default of 100
seconds.

We applied the tool to detector images that first underwent the
standard attitude correction process \citep{uchiyama:08a}.  We chose
circular regions with radius $\approx 2'$ that were visually centered
on the source image.  The tool then created a new attitude correction
file, which we applied with the \texttt{xiscoord} tool.  All of our
attitude correction files were created using solely the \xis1 images.

Pileup occurs in CCD detectors when two or more photons fall on the
same or neighboring pixels during the same readout frame, and
therefore are read as a single higher energy photon or are discarded
as a bad event \citep{davis:01a}. \texttt{pile\_estimate.sl} was run
after \texttt{aeattcor.sl}, as uncorrected blurring leads to an
\emph{underestimate} of the degree of pileup.

The tool first creates a lightcurve in three energy bands, using both
rate and counts. Here we used the default values of 0.5--1.5\,keV,
1.5--3\,keV, and 3--9\,keV for the energy bands and 32 seconds for the
time bins.  These data are then passed to the \texttt{vwhere}
\citep{noble:05a} filtering tool. The \texttt{vwhere} tool was used to
create the color-intensity diagrams shown in Fig.~\ref{fig:colors}),
and it was used to select times of approximately uniform rates and
colors (see \S\ref{sec:lightcurves}). These time selections were then
written to a filter file used for subsequent data extraction using the
\texttt{xselect} tool.

A box-car smoothed (3$\times$3 pixel bins) image is displayed using
the \texttt{ds9} tool, with the image rescaled to the estimated pile
up fraction.  Pile up fraction here is defined to be the ratio of
events lost via grade or energy migration to the events expected in
the absence of pileup.  Furthermore, the pile up fraction is based
upon the mean counts per 3$\times$3 pixel region per readout
frame. The image displays discrete steps that represent the minimum
pileup fraction in the displayed region, with the exception of the
highest displayed value. This value corresponds to the image maximum,
and is shown over regions at this maximum down to half way towards the
next lowest displayed value.

Using this image, we chose to exclude the central regions with pile up
fractions $\aproxgt 10\%$ (Fig.~\ref{fig:correction}).  Typically, this
was an $\approx 20''$ radius region, comprising $\approx 1/3$ of the detected
events.  The \texttt{pile\_estimate.sl} tool then gives an estimate of
the mean level of the remaining pile up fraction in the image, which
for these data was typically $<4\%$.

\clearpage

\begin{figure*}
\includegraphics[bb= 0 0 612 792,width=0.99\textwidth]{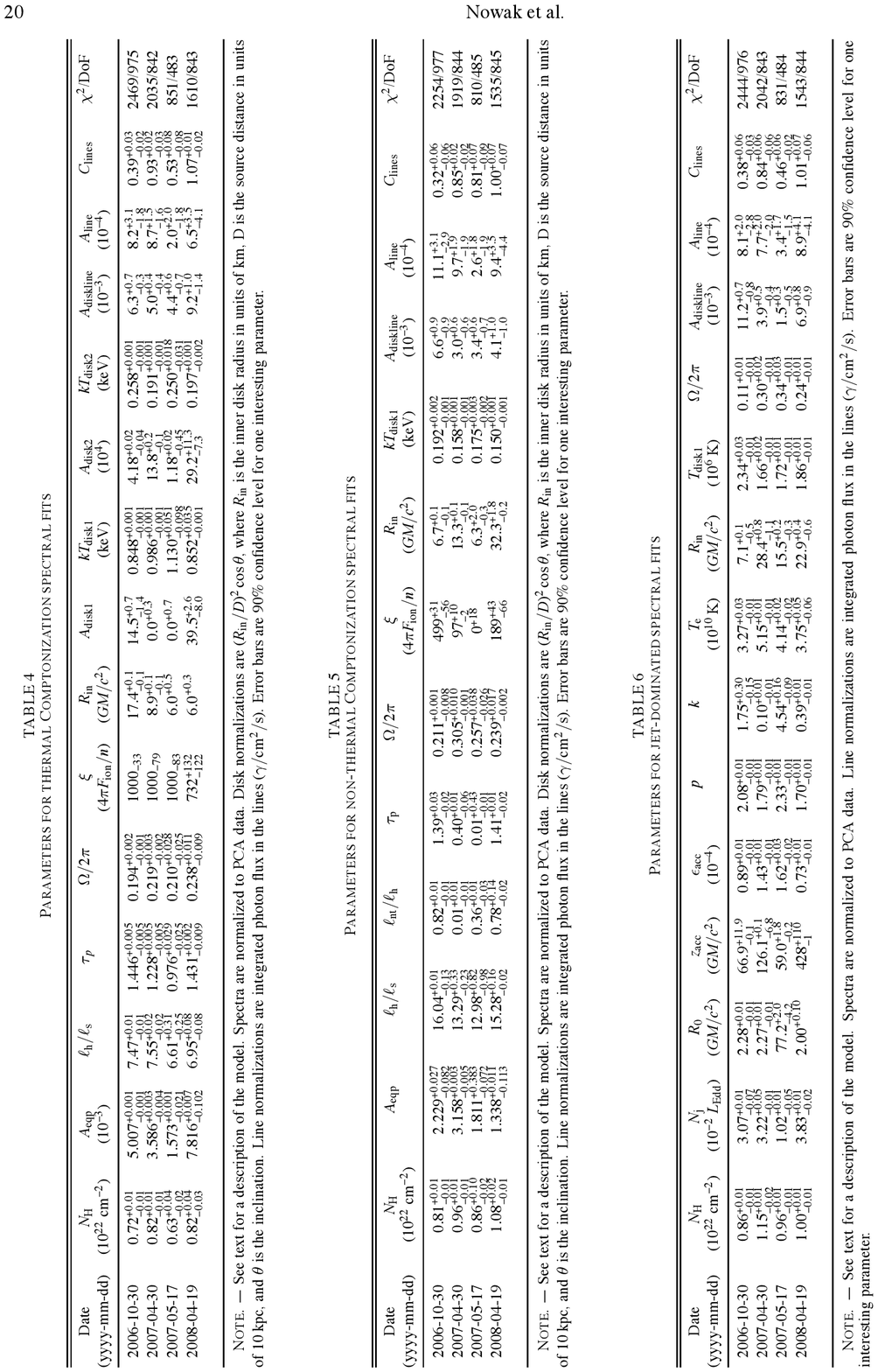}
\end{figure*}

\end{document}